\documentclass[11pt,a4paper]{article}
\pdfoutput=1
\usepackage{jcappub}
\usepackage{amsmath,amssymb,color,bm}
\usepackage{enumerate,xstring}

\def\be{\begin{equation}}
\def\ee{\end{equation}}
\def\bea{\begin{eqnarray}}
\def\eea{\end{eqnarray}}
\newcommand{\vs}{\nonumber\\}
\def\ba#1\ea{\begin{align}#1\end{align}}
\def\bg#1\eg{\begin{gather}#1\end{gather}}

\def\iMpch{\,h\,{\rm Mpc}^{-1}}

\newcommand{\refeq}[1]{Eq.~(\ref{eq:#1})}          
\newcommand{\refeqs}[2]{Eqs.~(\ref{eq:#1})--(\ref{eq:#2})}          
          
\newcommand{\reffig}[1]{Fig.~\ref{fig:#1}}          
          
\newcommand{\refsec}[1]{Sec.~\ref{sec:#1}}          
\newcommand{\refapp}[1]{Appendix~\ref{app:#1}}

\def\Plin{P_{\rm L}}
\def\Pnl{P_{\rm m}}
\newcommand{\fkone}{k\frac{\Plin^\prime(k)}{P_L(k)}}
\newcommand{\fktwo}{k^2\frac{\Plin^{\prime\prime}(k)}{P_L(k)}}
\newcommand{\fkonenl}{k\frac{P^\prime_m(k)}{P_m(k)}}
\newcommand{\fktwonl}{k^2\frac{P^{\prime\prime}_m(k)}{P_m(k)}}

\def\knl{k_\text{NL}}
\def\Ote{\hat{\varPi}}

\renewcommand{\v}[1]{\bm{#1}}

\newcommand{\vx}{\v{x}}

\newcommand{\vk}{\v{k}}
\newcommand{\vq}{\v{q}}
\newcommand{\vp}{\v{p}}

\newcommand{\<}{\langle}
\renewcommand{\>}{\rangle}

\renewcommand{\k}{\kappa}
\renewcommand{\d}{\delta}

\newcommand{\khat}{\hat{k}}

\def\xfl{\vx_{\rm fl}}

\def\be{\begin{equation}}
\def\ee{\end{equation}}
\def\ben{\begin{eqnarray}}
\def\een{\end{eqnarray}}
\def\ba{\begin{array}}
\def\ea{\end{array}}

\def\ba#1\ea{\begin{align}#1\end{align}}

\newcommand{\bq}{\begin{eqnarray}}
\newcommand{\eq}{\end{eqnarray}}
\newcommand{\bes}{\begin{subequations}}
\newcommand{\ees}{\end{subequations}}

\newcommand{\Om}{\Omega_m}

\def\R{\mathcal{R}}

\def\P{\mathcal{P}}

\def\O{\mathcal{O}}

\def\kunit{\:h\,{\rm Mpc}^{-1}}
\def\fii{F_2}
\def\fiii{F_3}
\def\cov{{\rm Cov}}

\def\khard{k_\text{hard}}
\def\ksoft{k_\text{soft}}

\makeatletter
\newlength{\apb@width}
\newcommand{\autoparbox}[2][c]{\settowidth{\apb@width}{#2}\parbox[#1]{\apb@width}{#2}}
\newcommand{\includegraphicsbox}[2][]{\autoparbox{\includegraphics[#1]{#2}}}
\makeatother

\newcommand{\lcdm}{$\Lambda${\rm CDM}}

\DeclareMathOperator{\tr}{tr}
\DeclareMathOperator{\Cov}{Cov}

\def\convD{\frac{\text{D}}{{\text{D}}\tau}}

\def\lapl{\nabla^2}

\newcommand{\comment}[1]{}

\def\cH{\mathcal{H}}
\def\Mpch{\,h^{-1}\,{\rm Mpc}}

\begin{document}

\title{Responses in Large-Scale Structure}

\author{Alexandre Barreira and}
\emailAdd{barreira@MPA-Garching.MPG.DE}

\author{Fabian Schmidt}
\emailAdd{fabians@MPA-Garching.MPG.DE}

\affiliation{Max-Planck-Institut f{\"u}r Astrophysik, Karl-Schwarzschild-Str. 1, 85741 Garching, Germany}

\abstract{
  We introduce a rigorous definition of general \emph{power-spectrum responses} as resummed vertices with two hard and $n$ soft momenta in cosmological perturbation theory. These responses measure the impact of long-wavelength perturbations on the local small-scale power spectrum. The kinematic structure of the responses (i.e., their angular dependence) can be decomposed unambiguously through a ``bias'' expansion of the local power spectrum, with a fixed number of physical \emph{response coefficients}, which are only a function of the hard wavenumber $k$. Further, the responses up to $n$-th order completely describe the $(n+2)$-point function in the squeezed limit, i.e. with two hard and $n$ soft modes, which one can use to derive the response coefficients. This generalizes previous results, which relate the angle-averaged squeezed limit to isotropic response coefficients. We derive the complete expression of first- and second-order responses at leading order in perturbation theory, and present extrapolations to nonlinear scales based on simulation measurements of the isotropic response coefficients. As an application, we use these results to predict the non-Gaussian part of the angle-averaged matter power spectrum covariance ${\rm Cov}^{\rm NG}_{\ell = 0}(k_1,k_2)$, in the limit where one of the modes, say $k_2$, is much smaller than the other. Without any free parameters, our model results are in very good agreement with simulations for $k_2 \lesssim 0.06\kunit$, and for any $k_1 \gtrsim 2 k_2$. The well-defined kinematic structure of the power spectrum response also permits a quick evaluation of the angular dependence of the covariance matrix. While we focus on the matter density field, the formalism presented here can be generalized to generic tracers such as galaxies.
  }


\date{\today}

\maketitle
\flushbottom

\section{Introduction}\label{sec:intro}

The large-scale distribution of matter in the Universe encodes a very rich set of observational imprints that can and have been used to test cosmological models. The standard way to describe its statistical properties is via $n$-point correlation functions of the matter density fluctuations field $\delta(\vx)$ \cite{Bernardeau/etal:2002}. The simplest such object is the $2$-point correlation function, $\xi(r)$, which measures the correlations of the density contrast in regions of the Universe separated by a distance $r$. Its Fourier counterpart is the matter power spectrum, 
\be
\<\delta(\vk)\delta(\vk')\> = (2\pi)^3\delta_D(\vk + \vk') \<\delta(\vk)\delta(\vk')\>' = (2\pi)^3\delta_D(\vk + \vk') P_m(k)\,.
\ee
Here and throughout, angle brackets denote an ensemble average and a prime on a correlator indicates that the overall momentum conserving $(2\pi)^3 \d_D(\vk_{\rm tot})$ factor is dropped. For an initially Gaussian distributed density field, and during the linear regime of structure formation, the power spectrum encodes all the statistical information of the matter field. Higher order $n$-point functions become important during the later stages of structure formation, when mode-coupling between different Fourier modes develops specific non-Gaussian signatures; or when the primordial density fluctuations are themselves non-Gaussian, as predicted by a range of inflation models \cite{2004PhR...402..103B}. Accurate theoretical predictions of these higher-order correlations are thus necessary to properly exploit the statistical information of the observational data. Even studies that rely solely on comparing the predicted matter power spectrum with observations, such as gravitational lensing, require a good understanding of higher-order correlations because the mode coupling adds an important contribution to the matter power spectrum covariance. This is described by a specific configuration of the matter trispectrum, the Fourier transform of the connected $4$-point correlation function \cite{1999ApJ...527....1S, 2013PhRvD..88f3537D}.

The complexity of modeling $n$-point functions increases rapidly with $n$. This holds for perturbation theory methods (see Ref.~\cite{Bernardeau/etal:2002} for a review) as well as measurements in N-body simulations. 
Perturbative approaches, including the effective field theory (EFT) approach 
\cite{baumann, 2012JHEP...09..082C} (see Ref.~\cite{porto:2016} for a review), are limited to the quasi-linear regime, i.e.,  wavenumbers $k \lesssim \knl$, where $\knl = 0.3\kunit$ is the nonlinear scale at redshift $z=0$. Numerical simulations of structure formation are currently the only available tool to accurately predict matter fluctuations in the nonlinear regime, but these predictions do not come without downsides. Simulations require significant amounts of computational resources, which makes it harder to obtain quick predictions, needed for instance to cover multidimensional spaces of cosmological models. Higher-order $(n>2)$-point correlations require significantly larger volumes to obtain converged results, as they live in higher-dimensional parameter spaces. Furthermore, estimators of higher-order correlations become themselves more computationally demanding. Attempts at precision measurements of the bispectrum $B(\vk_1, \vk_2, \vk_3)$ (3-point correlation function) illustrate these challenges \cite{2010MNRAS.406.1014S, 2015PhRvD..91d3530S, Lazanu/etal:15}.

 A simplification of the study of higher order correlation functions can be achieved by focusing on so-called {\it squeezed limit} configurations, i.e.,
 \be
 \<\delta(\vk)\delta(\vk')\delta(\vp_1)\delta(\vp_2)\cdots\delta(\vp_n)\>_c
 \,,
 \quad  \mbox{with} \quad p_i \ll k,k'\ (i=1,2,\cdots,n)\  \mbox{and}\  p_{12..n} \ll k,k'\,.
 \label{eq:sqcorr}
 \ee
Here, a subscript $_c$ indicates that we are considering only the connected part of the correlator, and we adopt a shorthand notation for the sum of vectors: $\vk_{12\cdots n} = \vk_1 + \vk_2 + \cdots \vk_n$. We shall also denote magnitudes of vectors as $k = |\vk|$.  
This squeezed-limit $(n+2)$-point function represents the coupling of $n$ long-wavelength (or soft) modes with two short-wavelength (or hard) modes. Symmetries of the large-scale structure provide strong constraints on this squeezed limit, a result that is known as ``consistency relations'' \cite{r1,r2,r3,r4,r5,r6,r7,r8,r9,r10,r11}. Further, Ref.~\cite{response} linked a particular angle-average of these squeezed configurations of the matter $(n+2)$-point function to the $n$-th order response $R_n(k)$ of the local matter power spectrum to an initial density perturbation (more precisely, for the case of $n$ superimposed spherically symmetric soft modes). These responses, which are a subset of more general \emph{response coefficients} defined below, can be measured accurately with separate universe simulations \cite{2011ApJS..194...46G, 2011JCAP...10..031B, 2016JCAP...09..007B, wagner/etal:2014}, which incorporate spherically symmetric long-wavelength perturbations by simulating curved Friedmann-Robertson-Walker cosmologies (see also Refs.~\cite{takada/hu:2013, li/hu/takada, lazeyras/etal, 2016arXiv161204360L, 2017arXiv170103375C} for further applications of the separate universe approach).

One of the main goals of this paper is to show how the relation between squeezed-limit $(n+2)$-point correlation functions and responses can be generalized beyond the special case of $n$ spherically symmetric perturbations, to cover the full shape of these correlation functions at leading order in $p_i/k$. This is related to the multipoint-propagator formalism \cite{bernardeau/etal:2008}, although the latter works at the level of the density field, not the power spectrum considered here. For any given value of $n$, we will see that there is a well-defined, finite set of response coefficients $R_O(k)$, which includes the above-mentioned isotropic response coefficient $R_n(k)$. One can think of these coefficients as describing the response of the local nonlinear matter power spectrum $P_m(\vk;\vx,\tau)$ measured around position $\vx$ at conformal time $\tau$ to the leading local gravitational observables. These observables include the density perturbation, tidal field, as well as convective time derivatives thereof. The $R_O(k)$ can therefore be regarded as the coefficients of a ``bias expansion'' of the local nonlinear matter power spectrum, in analogy with the expansion of galaxy bias (see Ref.~\cite{biasreview} for a review).

The response coefficients $R_O(k)$ are physical observables that can be measured in simulations, and one can use physical considerations to evaluate their magnitude and scale dependence (see for example the detailed discussions in Refs.~\cite{li/hu/takada,response}). We describe a procedure which uses perturbation theory results on large scales, together with the simulation measurements of the isotropic response coefficients presented in Ref.~\cite{response}, to make physically well-motivated estimates for the $R_O(k)$. Note that these responses, which quantify the effect of long-wavelength perturbations on the nonlinear gravitational evolution of the small-scale power spectrum, are to be distinguished from the response of the power spectrum to changes in the initial power spectrum, which were measured in Refs.~\cite{neyrinck/yang, nishimichi/bernardeau/taruya}.

The sequence of steps followed in this paper can be outlined as follows:

\begin{enumerate}[i.]
\item First, we rigorously define general \emph{power spectrum responses}, which we denote as $\R_n$ ($n=1,2,\cdots$) (\refsec{response}). These are functions of several angles and momentum ratios and describe the general (i.e., not angle-averaged) squeezed limit of ($n+2$)-point functions involving two hard and $n$ soft modes. The small-scale modes are allowed to be fully nonlinear. A diagrammatic representation of these {\it response-type} interactions shows how these responses can be used in more general perturbative calculations. 

\item We write the local power spectrum as a bias-like expansion of \emph{response coefficients} $R_O(k)$ that multiply all leading gravitational observables on which the local power spectrum can depend (namely, the density and tidal fields and their time derivatives) at any given order in perturbation theory (\refsec{biasexp}). In this way, the $\R_n$ are decomposed into a finite set of functions of $k$ only, multiplied by kernels that are uniquely determined by perturbation theory. This decomposition enormously simplifies the description of the responses $\R_n$. 
  
\item We then use these results to provide fully nonlinear predictions for $\R_1$ and $\R_2$, which are exact on large scales, but use a physically motivated extrapolation of simulation results on small scales. First, we derive the large-scale predictions for the $R_O(k)$ relevant for $\R_1$ and $\R_2$ by matching to the tree-level matter bispectrum and trispectrum, respectively (\refsec{derivation}). We then use the measured simulation results for the isotropic coefficients $R_1(k),\,R_2(k)$ presented in Ref.~\cite{response} to extrapolate all $R_O(k)$ to nonlinear scales, by employing a separation into ``growth'' and ``dilation'' effects (\refsec{extrapolation}).
\end{enumerate}

As an interesting first application of our framework, we use our description of $\R_2$ to predict the squeezed limit of the matter power spectrum covariance, as was already suggested by Ref.~\cite{bertolini1}. 
This corresponds to a special case of \refeq{sqcorr} with $n=2$, $\vk'=-\vk$, $\vp_2=-\vp_1$, and $p_1 = p_2 \ll \knl$ (\refsec{sqcov}). We shall see that this formalism, which combines analytical results with small-volume simulation measurements, allows one to match covariance matrix estimates based fully on numerical simulations to very good degree all the way up to $k = 1\kunit$ (the interested reader might have a quick look at Fig.~\ref{fig:sqcov} on p.~\pageref{fig:sqcov}). We also expand the squeezed covariance in a Legendre multipole expansion and briefly analyze its three non-vanishing moments (monopole, quadrupole and hexadecupole). In Sec.~\ref{sec:conc}, we summarize our main conclusions and outline possible future applications of the framework. Finally, Appendix \ref{app:feynman} lists the Feynman diagram conventions that we adopt in the main body of the paper; in Appendix \ref{app:tritree}, we illustrate explicitly the equivalence between $\R_2$ and the $4$-point connected correlation function in the squeezed limit; and in Appendix \ref{app:RLag}, we specify the distinction between Eulerian and Lagrangian definitions of power spectrum responses.

Throughout this paper, we always assume a flat $\Lambda{\rm CDM}$ cosmology with the following parameters (the same as those of the covariance matrix estimates from simulations in Ref.~\cite{blot2015}): $h = 0.72$, $\Omega_mh^2 = 0.1334$, $\Omega_bh^2 = 0.02258$, $n_s = 0.963$, $\sigma_8(z=0) = 0.801$.

\section{Power spectrum response: definition and connection to squeezed $(n+2)$-point functions}\label{sec:response}

In this section, we use diagrammatic representations of interactions (or mode-coupling) in cosmological perturbation theory to define the power spectrum responses, and show how they are directly related to certain squeezed limits of $(n+2)$-point correlation functions (see \refapp{feynman} for a summary of the Feynman rules). We begin with the simplest case of the first-order response, and then generalize to $n$-th order.

\subsection{First-order response}\label{sec:R1def}

Consider the nonlinear matter power spectrum, which is denoted as a ``blob'' propagator with two outgoing modes,
\be
\raisebox{-0.0cm}{\includegraphicsbox[scale=0.8]{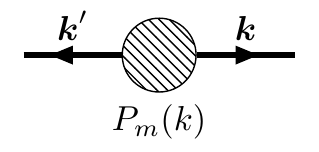}}
\equiv P_m(k,t)  \: (2\pi)^3 \d_D(\vk+\vk')\,.
\label{eq:Pmdef}
\ee
Throughout, arrows denote the direction of momentum as well as time.
Further, whenever it leads to no confusion, we suppress the time argument $t$ to shorten the notation. We will discuss aspects of the time dependence at the end of this section. 
The nonlinear power spectrum is a non-perturbative quantity, and the blob can be understood as resumming infinitely many perturbative contributions including counterterms. 
The linear power spectrum on the other hand will be denoted with a dot in our notation.

Now, consider the following, also non-perturbative, $3$-point interaction vertex, with two outgoing hard (high-momentum, or wavenumber) Fourier modes, and one ingoing soft (low-momentum) mode, all defined at a fixed time $t$: 
\be
\lim_{p \to 0} \left(
\raisebox{-0.0cm}{\includegraphicsbox[scale=0.8]{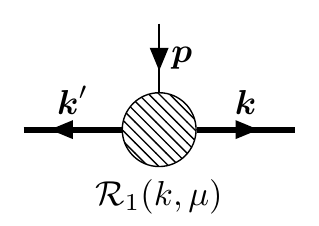}}
\right)
\equiv \frac12 \R_1(k; \mu_{\vk,\vp}; t) P_m(k, t) \: (2\pi)^3 \d_D(\vk+\vk'-\vp)\, .
\label{eq:fR1def}
\ee
where $\mu_{\vk,\vp} = \vk\cdot\vp/(k p)$. 
Here and throughout, thick and thin lines denote hard and soft lines, respectively. Further, the notation $\lim_{p\to 0}$ is not to be understood as mathematical limit, but signifies that only the leading contribution in this limit, i.e. the lowest power of $p$, is kept. 
The resummed vertex in \refeq{fR1def} defines the first-order power spectrum \emph{response} $\R_1$. 
The meaning of $\R_1$ can be elucidated as follows. In the limit $p/k\to 0$, we have $\vk' \approx -\vk$. That is, up to corrections suppressed by $p/k$, the hard modes are in the same configuration as in the case of the nonlinear matter power spectrum \refeq{Pmdef}. 
The comparison between Eqs.~(\ref{eq:fR1def}) and (\ref{eq:Pmdef}) then provides justification to call $\R_1(k; \mu_{\vk,\vp})$ a power spectrum response, in that it describes the impact of the  linearly evolved soft mode $\d^{(1)}(\vp)$ on the nonlinear power spectrum. Note that we do not require that $\vk,\vk'$ be in the perturbative regime, i.e., they are allowed to be fully nonlinear. On the other hand, in addition to having to be much smaller than $k$, $p$ is also assumed to be in the perturbative regime, i.e. $p \ll \knl$. One might wonder why we have not allowed for $\R_1$ to depend on $p$. The reason is that in the low-$p$ limit, a long-wavelength perturbation modulating the power spectrum appears as a uniform (spatially constant) contribution to the density and tidal field. Any dependence on the wavelength of the soft mode enters only at order $(p/k)^2$ or $(p/\knl)^2$ (whichever is larger). The same will correspondingly hold for higher-order responses. This will be explicitly justified in the next section.

The vertex corresponding to $\R_1$ has three lines, suggesting that its leading contribution will be at the three-point function (bispectrum) level. Indeed, we can obtain its contribution to the (equal-time) bispectrum by attaching a soft power spectrum as in 
\be\label{eq:bispecR1}
\lim_{p \to 0} \left(
\raisebox{-0.0cm}{\includegraphicsbox[scale=0.8]{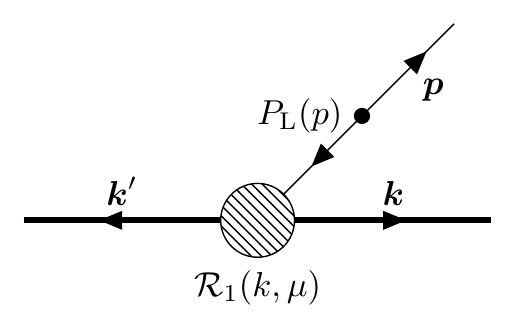}}
+ ( \vk \leftrightarrow \vk' )
\right)
= \R_1(k; \mu_{\vk,\vp}) P_m(k) \Plin(p)\: (2\pi)^3 \d_D(\vk+\vk'+\vp)\,,
\ee
where the black dot represents the linear power spectrum $\Plin(p)$ (see \refapp{feynman}).  This constitutes the dominant contribution to the matter bispectrum, $B(\vk, \vk', \vp)$ in the \emph{squeezed limit} $p/k \ll 1$.  
To demonstrate this, we can work at tree level in perturbation theory, in which case we have that 
\bq\label{eq:bispectree}
B^{\rm tree}(\vk, \vk', \vp)= 2\Big[\fii(\vk,\vp)\Plin(k) + (\vk \leftrightarrow \vk') \Big]\Plin(p) + 2 \fii(\vk, \vk')\Plin(k)\Plin(k'),
\eq
where $\fii$ is the symmetrized second order perturbation theory kernel (see Eq.~(\ref{eq:delta2}) below). The two terms in brackets in Eq.~(\ref{eq:bispectree}) correspond to the diagram of Eq.~(\ref{eq:bispecR1}) at tree level.\footnote{At tree level in perturbation theory, the blob in Eq.~(\ref{eq:bispecR1}) turns into an $F_2$ vertex with a propagator for momentum $\vk$ (and similarly for $\vk'$ to account for the $\vk \leftrightarrow \vk'$ permutation).}  The second term in Eq.~(\ref{eq:bispectree}), $\propto \Plin(k)\Plin(k')$, corresponds to a diagram where two hard ingoing modes form a soft outgoing mode, and it is not classifiable as a power spectrum response.  However, in the squeezed limit that we are considering, this interaction is suppressed by $\fii(\vk, \vk') \propto \left(p/k\right)^2$. This is required by mass and momentum conservation: by definition, small-scale perturbations can only redistribute the mass within a region that is much smaller than the scale $1/p$ of the long mode. This means that fluid momentum generated by the coupling of small-scale modes is of order $i\vp/k^2$, which leads to a contribution to the large-scale density suppressed by $(p/k)^2$ (see Appendix~B of Ref.~\cite{abolhasani/mirbabayi/pajer:2016} for a more detailed discussion).  Beyond this, there is a further suppression by $\Plin(k)/\Plin(p)$ which, given $p \ll k$, is much less than 1 for $p \gtrsim 10^{-2} \iMpch$ due to the shape of the power spectrum.

As shown in more detail in Sec.~\ref{sec:derivation}, the tree-level prediction for $\R_1$ can be read off from equating the first two contributions in \refeq{bispectree} to \refeq{bispecR1}:
\bq\label{eq:R1tree}
\R_1^\text{tree}(k;\mu) &=& \frac{47}{21} - \frac{1}{3}\fkone
+ \left(\frac{8}{7} - \fkone\right) \left(\mu^2-\frac13\right)
\,.
\eq
Note that indeed, $\R_1^\text{tree}$ is independent of $p$, as it should be. Moreover, the dependence on $\mu$ simply consists of a monopole and a quadrupole. We will see below that the fully nonlinear response $\R_1$ retains a very similar structure, and is completely described by two functions of $k$, which reduce to \refeq{R1tree} on large scales.
The physical interpretation of the angular structure (i.e., the $\mu$ dependence) is addressed in the next section.  
Beyond tree level, we can write for the squeezed-limit bispectrum,
\bq\label{eq:Bsq}
\lim_{p \to 0} B(\vk,\vk',\vp) = \lim_{p \to 0} \< \d(\vk) \d(\vk') \d(\vp)  \>'_{c} =  \R_1(k,\mu_{\vk,\vp}) P_m(k) \Plin(p)\,,
\eq
where corrections away from the limit $p\to 0$ are suppressed by $(p/k)^2$ and $(p/\knl)^2$.

\subsection{Generalization to $n$-th order responses}\label{sec:genn}

Having gained intuition with the simpler first-order case, we now provide the definition of the general $n$-th order response $\R_n$. It is defined analogously to $\R_1$ in \refeq{fR1def} as 
\ba
\lim_{\{p_a\} \to 0} \left(
\raisebox{-0.0cm}{\includegraphicsbox[scale=0.8]{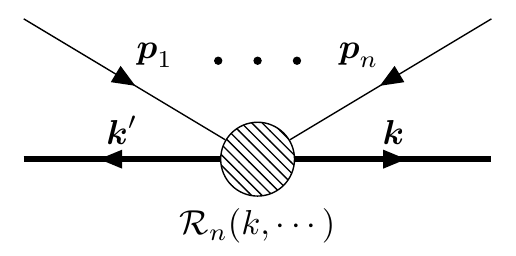}}
\right)
=\:& \frac12 \R_n(k;\, \{\mu_{\vk,\vp_a}\},\, \{\mu_{\vp_a,\vp_b}\},\, \{p_a/p_b\}) P_m(k) \vs[-15pt]
&\times (2\pi)^3 \d_D(\vk+\vk'- \vp_{1\cdots n})\,,
\label{eq:Rndef}
\ea
which corresponds to the modulation of the nonlinear matter power spectrum $P_m(k)$ by $n$ linearly evolved long-wavelength modes $\vp_1,\,\cdots,\,\vp_n$. Here and throughout, the notation $\lim_{\{p_a\}\to 0}$ implies that only the lowest powers in all of the $p_a$ are kept. Specifically, the leading correction to \refeq{Rndef} for finite soft momenta is suppressed by
\be
\max\left\{\frac{p^2}{k^2},\,\frac{p^2}{\knl^2}\right\}
\quad\mbox{where}\quad p \equiv \max \{ p_a \}_{a=1,\cdots n} \,.
\ee
As we will see below in concrete examples, all singular terms in the $p_a\to 0$ limit cancel in \refeq{Rndef}, so that the limit can be taken in any order. 
Note that, in addition to $k$,  $\R_n$ depends on $n(n+1)/2$ cosines of wavenumbers, as well as on $n(n-1)/2$ relative magnitudes of soft momenta, adding up to a total of $n^2+1$ arguments including $k$. Recall that $\R_n$ does not depend on the overall scale of the $p_a$ in the limit $p_a/k\to 0$. Analogously to the relations in Eqs.~(\ref{eq:bispecR1}) and (\ref{eq:Bsq}) between the first-order response and the $3$-point correlation function, the $n$-th order response contributes to the connected
$(n+2)$-point function $\<\d(\vk)\d(\vk')\d(\vp_1)\cdots \d(\vp_n)\>_c$ in the kinematic regime where the modes $\vp_1,\vp_2, \cdots \vp_n$ are soft (the condition that $|\vp_{1\cdots n}| \ll k$ must also hold), through the following diagram: 
\ba
\lim_{\{p_a\}\to 0} \left(
\raisebox{-0.0cm}{\includegraphicsbox[scale=0.8]{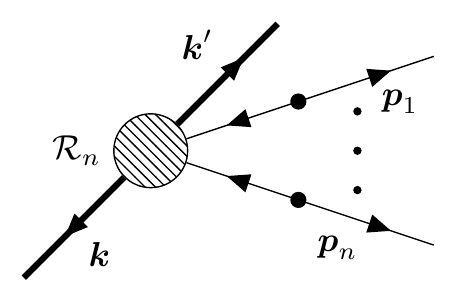}}
+ (\text{perm.}) \right)& = \<\d(\vk)\d(\vk')\d(\vp_1)\cdots \d(\vp_n)\>_{c, \R_n}
\vs
= n!\, \R_n(k;\, \{\mu_{\vk,\vp_a}\},\, \{\mu_{\vp_a,\vp_b}\},\, \{p_a/p_b\})
P_m(k) &\left[\prod_{a=1}^n\Plin(p_a)\right] \: (2\pi)^3 \d_D(\vk+\vk'+\vp_{1\cdots n})\,.
\label{eq:sqnpt}
\ea
The factor $n!$ arises from the permutations among the $\vp_a$. The subscript $\R_n$ on the correlator indicates that we are considering only the contributions that involve $\R_n$, and the squeezed limit is implicitly assumed.
In addition to \refeq{sqnpt}, there are two other types of terms that contribute to the $(n+2)$-point function in this limit: 
\begin{enumerate}
\item Contributions of the form
\be
\propto P_m(k)\underbrace{\Plin(p_a)\Plin(p_b)\Plin(|\vp_{ab}|) \cdots \Plin(|\vp_{c\cdots d}|)}_{n\:\text{soft power spectra}}.
\ee
These diagrams can be broken down into vertices involving only soft lines (which can be treated in standard perturbation theory) that interact with the nonlinear power spectrum in a response vertex of lower-order $\R_m$ ($m<n$ is the number of soft lines that attach to the $\R_m$ response vertex, and these must include at least one whose momentum is the sum of two or more soft momenta). A schematic example of such a term is
\ba
&\raisebox{-0.0cm}{\includegraphicsbox[scale=0.8]{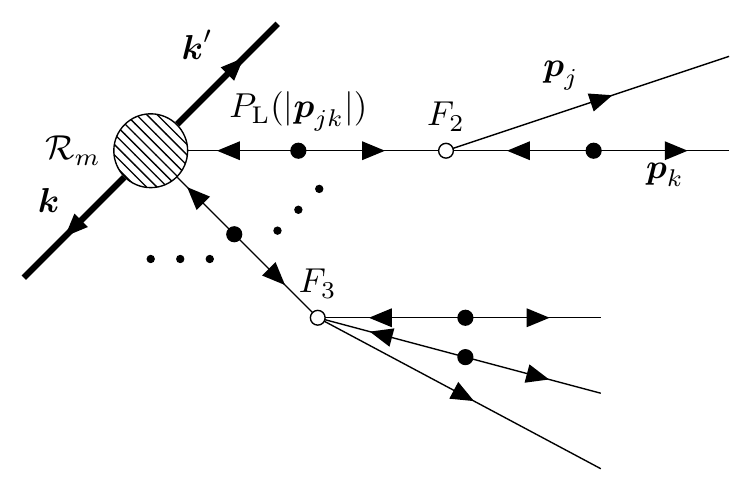}}.
\label{eq:sqnpt-lo}
\ea
Explicitly, at $n=2$, there is a single such contribution given by (see Appendix \ref{app:tritree}) 
\ba
&\raisebox{-0.0cm}{\includegraphicsbox[scale=0.8]{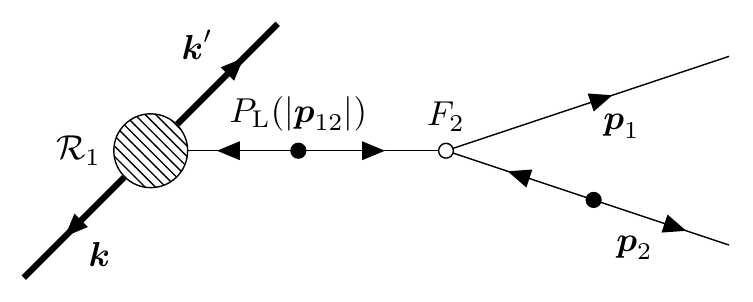}}
+ (\mbox{perm.}) \vs
&\hspace*{2cm} = \R_1(k; \mu_{\vk,\vp_{12}}) P_m(k) \left[ 2 F_2(-\vp_{12},\vp_2) \Plin(|\vp_{12}|) \Plin(p_2)+ (\vp_1 \leftrightarrow \vp_2) \right] \vs
&\hspace*{2.5cm} \times (2\pi)^3 \d_D(\vk+\vk'+\vp_{12}) \,.
\label{eq:sqTR1}
\ea

\item Contributions of the form
\ba
&\propto \Plin(k) \Plin(k') \Plin(p_2)\cdots\Plin(p_{n})\,,\vs
&\propto \Plin(k)\Plin(|\vk+\vp_a|) \Plin(|\vk+\vp_{ij}|) \Plin(p_3)\cdots\Plin(p_{n})\,,\quad\mbox{and so on,}
\ea
where we have only written tree-level contributions. These terms are characterized by two or more hard power spectra (and correspondingly $n-1$ or less soft power spectra). They necessarily result from two or more ingoing hard modes combining to soft modes, and are the generalization of the squeezed-bispectrum contribution $\propto \Plin(k)\Plin(k')$ in \refeq{bispectree}. Therefore, they do not correspond to response vertices. However, as discussed above, in the squeezed regime, they are highly suppressed both by mass-momentum conservation and by the shape of the matter power spectrum. If $n_h$ is the number of hard power spectra in a given contribution, they are suppressed by $(p/k)^{2(n_h-1)} [\Plin(k)/\Plin(p)]^{n_h-1}$, where again $p \equiv \max\{p_1, p_2, \cdots, p_n\}$. On the other hand, loop corrections are instead suppressed by factors of $(p/\knl)^{3+n_P}$, where $n_P = d\ln\Plin/d\ln k$ is the linear power spectrum index, evaluated at a scale of order $\knl$ (to be distinguished from the primordial spectral index $n_s$).

\end{enumerate}

To summarize, the leading contributions to the squeezed-limit $(n+2)$-point function with two hard and $n$ soft modes are completely described by the responses $\R_m$ with $m\leq n$. This contains the contribution of Eq.~(\ref{eq:sqnpt}), proportional to $\R_n$, as well as those of point 1 above, which are given in terms of $\R_m$, $1 \leq m < n$. We stress that this holds for fully nonlinear hard modes. The only restriction is that $p \ll {\rm min}\{ k,k',\knl\}$. In Appendix \ref{app:tritree}, we illustrate this in detail by listing all the terms for $n=2$ at tree level.

\bigskip

In the discussion presented thus far, the responses $\R_n$ are vertex interactions that depend, in addition to $k$, also on the orientations and relative magnitudes of the soft modes, for a total of $n^2+1$ arguments. One may therefore argue that, at least at first sight, the problem of calculating the squeezed limit of correlation functions has not been significantly reduced. Crucially however, as we will see in the next section, there is a well-defined decomposition of the $\R_n$ into a finite number of \emph{response coefficients} $R_O(k)$ which are only a function of $k$, multiplied by specific kinematic shapes, i.e. functions of the angles and relative magnitudes of the soft modes. For $\R_1$, there are two such coefficients and kinematic shapes, while for $\R_2$, there are six. Thus, instead of being a free function of five arguments, we will see that $\R_2$ is completely determined by six functions of $k$ only, which represents a significant reduction in functional freedom. These discussions are the subject of the next section.

\bigskip

Before continuing however, for completeness, we restore the time dependences in the relation between squeezed $(n+2)$-point functions and $\R_n$. Specifically, the responses describe the contributions to the \emph{unequal} time $(n+2)$-point functions in the limit $\{p_a\}\to 0$ as
\ba
& \<\d(\vk,t)\d(\vk',t)\d(\vp_1,t_1)\cdots \d(\vp_n,t_n)\>'_{c, \R_n} \vs
& \qquad = n!\, \R_n\left(k, t, \{t_a\};\, \{\mu_{\vk,\vp_a}\},\, \{\mu_{\vp_a,\vp_b}\},\, \{p_a/p_b\}\right)
P_m(k,t) \prod_{a=1}^n\Plin(p_a,t_a)\,,
\label{eq:sqnptT}
\ea
where, we recall, the subscript $\R_n$ indicates that only those contributions that are captured by $\R_n$ are considered, and the squeezed limit is implied.  
At tree level, i.e.~to zeroth order in $k/\knl$, the $\R_n$ are directly related to the perturbation theory kernels, and hence independent of time in an Einstein-de Sitter (EdS) universe. This also holds in \lcdm\ to better than percent-level accuracy. Beyond tree level, the time independence no longer holds. In the EdS limit however, this time dependence only appears in the response coefficients $R_O(k)$, which are multiplied by time-independent kernels. Thus, to percent-level accuracy, we can write
\be
\R_n\left(k, t, \{t_a\};\, \{\mu_{\vk,\vp_a}\},\, \{\mu_{\vp_a,\vp_b}\},\, \{p_a/p_b\}\right)
= \R_n\left(k, t;\, \{\mu_{\vk,\vp_a}\},\, \{\mu_{\vp_a,\vp_b}\},\, \{p_a/p_b\}\right)\,.
\ee
We stress that the response approach does not rely on the EdS approximation, and can be analogously performed using the exact \lcdm\ expansion history as well.

It is also worth noting that additional terms enhanced by $k/p_a$ appear in the squeezed limit of unequal-time $(n+2)$-point functions, which are induced by the displacement of the small-scale modes by the large-scale modes \cite{peloso/pietroni} (note that the terms derived in Ref.~\cite{peloso/pietroni} for the bispectrum appear only if the hard modes are evaluated at different times, whereas the response vertices always describe two hard modes at equal time). It can be shown that these contributions are also described by responses, and are of the type of \refeq{sqnpt-lo}, i.e., they involve $\R_m$ with $m<n$.

\section{Decomposition into response coefficients}\label{sec:biasexp}

We now show how the responses $\R_n$ can be decomposed into a set of $k$-dependent 
coefficients multiplying distinct kinematic shapes. For this, we first 
derive how the $\R_n$ measure the response, or modulation, of the matter power spectrum in the presence of $n$ long-wavelength, linearly evolved density fluctuations. This can be formulated precisely by defining the response as the functional derivative of the \emph{local} nonlinear power spectrum with respect to~the amplitude of the long-wavelength density perturbations, $\delta^{(1)}({\vp_a})$ \cite{response, bertolini1}
\bq\label{eq:respderi}
\R_n\big(k; \{\mu_{\vk,\vp_a}\}, \{\mu_{\vp_a,\vp_b}\}, \{p_a/p_b\} \big) = \frac{1}{n!\, P_m(k)}\frac{{\rm d}^n P_m(\vk | \delta^{(1)}({\vp_1}), \cdots, \delta^{(1)}({\vp_n}))}{{\rm d}\delta^{(1)}({\vp_1}) \cdots {\rm d}\delta^{(1)}({\vp_n})}\Big|_{\d^{(1)}(\vp_a)=0}\,. \nonumber \\
\eq
Here, we implicitly require that $p_a \ll k$, and $p_a \ll \knl$, since the notion of a local power spectrum is only meaningful in this limit. The meaning of $P_m$ in Eq.~(\ref{eq:respderi}) can be rigorously understood as follows. Let us consider a region of size $L$, with $1/k\ll L \ll 1/p$ and $L \gg 1/\knl$, and measure the Fourier modes $\d(\vk)$ within this volume. Since the large-scale modes can be approximated as constant over this region, it corresponds to a homogeneous (though not necessarily isotropic) patch of space, so that the local, anisotropic power spectrum of small-scale modes in the region of size $L$, $\< \d(\vk)\d(\vk')\>_L$, is meaningful.

In order to derive the functional derivative in \refeq{respderi}, we treat the local, small-scale power spectrum as a particular case of a biased tracer (see Sec.~2 of Ref.~\cite{biasreview} for a detailed introduction). For this, it is convenient to work in a mixed real/Fourier-space representation. While the small-scale modes are described in Fourier space (within the region of size $L$ as described above), we treat the large-scale modes in real space. In the large-scale bias expansion, we have to include all local gravitational observables that can be constructed out of the long-wavelength modes. As shown in Refs.~\cite{senatore:2014,MSZ}, these consist of all nonlinear local combinations of the tidal tensor $\partial_i\partial_j\Phi(\xfl[\tau],\tau)$, as well as its convective time derivatives. Here, $\xfl(\tau)$ denotes the fluid trajectory leading to a given Eulerian position $\vx$ at time $\tau$, while convective time derivatives are defined as $D/D\tau = \partial/\partial\tau + v^i\partial/\partial x^i$, where $v^i = dx_{\rm fl}^i/d\tau$ is the velocity of the matter fluid. Reference \cite{MSZ} derived a compact set of complete, linearly independent terms in the bias expansion at any order, on which we will rely in our considerations below. 

To start with, we write the local, anisotropic power spectrum within a region of size $L$ around position $\vx$ as a sum of operators $(\hat k^i \hat k^j \cdots)O_{ij\cdots}(\vx)$ describing the local gravitational observables corresponding to the long-wavelength modes, multiplied by ``bias,'' or response coefficients $R_O(k)$:
\be
\frac{P_m(\vk|\vx)}{P_m(k)}-1 = \sum_O R_O(k) \left(\hat k^i \hat k^j \cdots \right) O_{ij\cdots}(\vx)\,,
\label{eq:Pkexp}
\ee
where we have continued to suppress the time argument (recall that $R_O \equiv R_O(k,t)$ in general). The notation $\left(\hat k^i \hat k^j \cdots \right)$ stands for a product of unit vectors that are contracted with $O_{ij\cdots}$, and which effectively take into account the preferred directions induced in the local patch by the long-wavelength modes. Let us discuss a few noteworthy aspects of the expansion of \refeq{Pkexp}. As before, we define $p \equiv \max\{p_1, p_2,\cdots, p_n\}$. 
\begin{enumerate}
\item The large-scale modes provide preferred directions, so that the growth of small-scale perturbations within the region considered depends on the angle of the small-scale modes with these directions. We have therefore allowed the local power spectrum to be direction-dependent.  Note however that $P_m(-\vk|\vx) = P_m(\vk|\vx)$ still has to hold, since the density field is real. Contributions of the form $\hat k^i O_i$, or in fact any term involving odd powers of $\hat{\vk}$, are therefore forbidden.  Given these constraints and the results of the bias expansion of general tracers, there is a well-defined set of operators appearing in the expansion of \refeq{Pkexp} at any given order in perturbations.
\item By definition, the operators $O_{ij\cdots}$ describe the modulation of the local power spectrum within the region of size $L \ll 1/p$. They should therefore be interpreted as coarse-grained on the scale $L$, although we do not emphasize this in the notation.  Moreover, in analogy to the bias expansion for galaxy or halo overdensities \cite{mcdonald:2006,assassi/etal}, the operators $O_{ij\cdots}$ are to be considered as renormalized.  This implies in particular that $\<O_{ij\cdots}\>=0$.
\item In real space and evaluated along a fixed fluid trajectory, each of the leading operators $O_{ij\cdots}$ is given by a combination of terms which each involve exactly two spatial derivatives acting on the gravitational potential $\Phi$ \cite{senatore:2014,MSZ}. Note that $O_{ij\cdots}$ does in general contain terms that are nonlocal in $\partial_i\partial_j\Phi$, such as, at second order, $(\partial_i\partial_j/\nabla^2)(\nabla^2\Phi)^2$. When evaluating these operators in perturbation theory at a fixed Eulerian position $(\vx,t)$, we also obtain terms that correspond to the displacement of the fluid trajectory from the Lagrangian position. These terms are of the form $s^k \partial_k O_{ij \cdots}$, where $\v{s}(\vq,t) = \xfl(\vq,t)-\vq$ is the Lagrangian displacement and $\vq$ is the Lagrangian coordinate of the fluid trajectory.  
\item Beyond the leading local gravitational observables mentioned above, there are also contributions to the bias expansion that have more than two spatial derivatives of $\Phi$ in \refeq{Pkexp}. Each additional derivative yields a power of $p_a \leq p$. Since odd powers of $k$ are ruled out by the constraint that the density field is real, the leading such higher-derivative term is of order $p^2$. The coefficients of these higher-derivative terms contain a spatial scale which describes the size of the region ``probed'' by the tracer. In our case, the tracer is the matter power spectrum $P_m(\vk|\vx)$, which means that the spatial scale has to be at least $1/k$. Furthermore, even very small-scale modes $k \gg \knl$ probe scales that are of order the nonlinear scale $2\pi/\knl \sim 20 \Mpch$, which corresponds to the typical distance traveled by dark matter particles over the course of structure formation. In summary, the leading higher-derivative contributions to \refeq{Pkexp} are expected to scale as
  \be
\max\left(\frac{p^2}{k^2},\,\frac{p^2}{\knl^2}\right)\,,
\ee
and are therefore small. In fact, these higher-derivative terms are of the same order as the terms in squeezed-limit $(n+2)$-point functions that are not captured by the response definition of \refeq{fR1def} and \refeq{Rndef}. We therefore do not include them throughout. In principle, it is possible to do so, thereby allowing the response approach to recover as well subleading contributions to squeezed-limit $(n+2)$-point functions, but we do not do this here.
\item At any given order $m$ in standard perturbation theory, the operators $O_{ij\cdots}$ contracted with $\hat k^i\hat k^j\cdots$ can be written as
  \be
 \left(\hat k^i\hat k^j\cdots\right)  O_{ij\cdots}^{(m)}(\vx) = \left[\prod_{a=1}^m \int_{\vp_a}\right] e^{i\vx\cdot\vp_{12\cdots m}}  \mathcal{K}^{(m)}_{O}(\{\hat{\vp}_b\},\, \{p_b/p_c\})\: \d^{(1)}(\vp_1)\cdots \d^{(1)}(\vp_m)\,,
  \label{eq:Okernel}
  \ee
  where $\mathcal{K}^{(m)}_O$ is a kernel that is specific to the $m$-th order contribution in perturbation theory to the operator $O$. Similarly to the perturbation theory kernels $F_n$, we assume without loss of generality that $\mathcal{K}^{(m)}_O$ is fully symmetrized in its arguments. The Fourier-space kernel only involves the unit vectors $\hat{\vp}_a$ as well as relative magnitudes $p_b/p_c$ of the soft modes. Even though, as mentioned above, nonlocal terms are present in the $O_{ij\cdots}$ in real space, the Fourier representation is an algebraic function of these quantities. We will encounter concrete examples illustrating \refeq{Okernel} and the properties of $\mathcal{K}_O$ below. In \refeq{Okernel}, as throughout, we have implicitly assumed that all quantities are evaluated at a fixed time $t$. In general, $\mathcal{K}_O$ is a function of $t$ as well, although it is independent of time in the EdS approximation, which we will assume throughout. If better than percent-level precision in \lcdm\ is desired, then this time dependence can be straightforwardly incorporated.
\end{enumerate}

We are now in a position to decompose the responses. At fixed order $n$ in perturbation theory, \refeq{Pkexp} is given by
\be
\frac{P_m(\vk|\vx)}{P_m(k)}-1 \stackrel{\text{$n$-th order}}{=} \sum_{O} R_O(k) \left(\hat k^i \hat k^j \cdots \right) O_{ij\cdots}^{(n)}(\vx)\,,
\label{eq:Pkexp2}
\ee
where the sum runs over all operators which start at $n$-th \emph{or lower} order in perturbation theory (we will see concrete examples below). Crucially, the coefficients $R_O(k)$ are independent of the order $n$ in perturbation theory that is employed to describe the soft modes $\vp_a$ that make up the operators $O_{ij\cdots}^{(n)}$.
After inserting \refeq{Okernel} into \refeq{Pkexp2}, and the resulting equation into \refeq{respderi}, we can write the response $\R_n$ as a sum over kernels multiplied by coefficients that only depend on $k$:
\ba
\R_n\big(k, t; \{\mu_{\vk,\vp_a}\}, \{\mu_{\vp_a,\vp_b}\}, \{p_a/p_b\} \big) =\:&
\sum_O R_O(k,t) \mathcal{K}^{(n)}_{O}(\{\hat{\vp}_a\}, \{p_a/p_b\})\,,
\ea
where the sum runs over the same operators as in \refeq{Pkexp2}. This exercise therefore organizes the cumbersome multi-dimensional dependencies of $\R_n$ into a simpler linear combination of the response coefficients $R_O(k,t)$, which depend only on the scale $k$ and time $t$, and that multiply the kernels $\mathcal{K}^{(n)}_O$, which unambiguously fix the possible angular and configuration dependencies. We stress that while the kernels $\mathcal{K}^{(n)}_O$ are independent of time in the commonly used EdS-approximated standard perturbation theory approach, as mentioned above, the $R_O(k,t)$ always depend on time, as they capture fully nonlinear structure formation.

Before continuing, we note that the $R_O$ can be seen as ``Eulerian'' response coefficients, as we have expanded the matter power spectrum in terms of operators at the final time. In \refapp{RLag}, we introduce a slightly different definition of ``Lagrangian'' response coefficients $R_O^L$, which are a generalization of those introduced in Ref.~\cite{response}, and analogous to the Lagrangian bias coefficients of dark matter halos. These correspond to a different basis in the same vector space, i.e., at any given order $n$, the Lagrangian $R_O^L$ can be expressed in terms of the Eulerian $R_O$, and vice versa. Crucially however, the expression for the full response $\R_n$ is independent of the choice of Eulerian or Lagrangian, or any other equivalent basis.

\subsection{General power spectrum expansion}
\label{sec:genexp}

As mentioned above, the expansion of \refeq{Pkexp} can be constructed out of 
$\partial_{i}\partial_{j}\Phi(\vx,\tau)$ and its convective time 
derivatives.  We now provide a recipe to write down this expansion by following Ref.~\cite{MSZ}, who defined\footnote{Note that the prefactor $2/(3\Om\cH^2)$ is absorbed into the definition of $\Phi$ there.}
\be
\Pi^{[1]}_{ij}(\vx,\tau) = \frac{2}{3\Om\cH^2} \partial_{i}\partial_{j}\Phi(\vx,\tau)
=
K_{ij}(\vx,\tau) + \frac13 \delta_{ij}\d(\vx,\tau)\,,
\label{eq:hatPi}
\ee
where $\Omega_m(\tau)$ is the ratio of matter density to critical density and $\cH = a^{-1}{\rm d}a/{\rm d}\tau$ is the conformal Hubble rate. $\Pi^{[1]}$ contains the density perturbation and the tidal field via $\d = \tr[ \Pi^{[1]} ]$ and
\be\label{eq:Del}
K_{ij} \equiv \Pi^{[1]}_{ij} - \frac13 \d_{ij} \tr [\Pi^{[1]} ]
= \left(\frac{\partial_i\partial_j}{\lapl} - \frac13 \d_{ij} \right) \d\,.
\ee
The superscript $[1]$, which is to be distinguished from $(1)$, refers to the
fact that $\Pi^{[1]}$ \emph{starts} at first order in perturbation theory,
but contains higher order terms as well. 
We then recursively define higher-order tensors $\Pi^{[n]}$ 
by taking convective time derivatives as (see Ref.~\cite{MSZ} for more details)
\be
\Pi^{[n]}_{ij} = \frac{1}{(n-1)!} \left[(\cH f)^{-1}\convD \Pi^{[n-1]}_{ij} - (n-1) \Pi^{[n-1]}_{ij}\right]\,,
\label{eq:Pindef}
\ee
where $f = d\ln D/d\ln a$ is the linear growth rate and $D(\tau)$ is the linear growth factor. As shown in Ref.~\cite{MSZ} for the case of the galaxy number density, the complete bias expansion consists of all scalar combinations of the $\Pi^{[n]}_{ij}$ that are relevant at any given order.  However, the operators $\tr[\Pi^{[n]}]$ with $n>1$ are degenerate with lower-order operators, since they are completely determined by the equations of motion for matter. For this reason, they do not need to be included in the bias expansion.

We now wish to generalize this expansion to the local matter power spectrum. Thus, we include all terms that have an even number of indices to contract with $\hat k^i \hat k^j \cdots$, i.e. 2-tensors, 4-tensors, and so on, again excluding $\tr[\Pi^{[n]}]$ with $n>1$. Up to second order, we have
\bea
\label{eq:EulBasis} {\rm 1^{st}\ order} \ && \ \d_{ij} \tr[\Pi^{[1]}]\,,\  \Pi^{[1]}_{ij} \\[3pt] 
\label{eq:EulBasis2}{\rm 2^{nd}\ order } \ && \ \d_{ij}\d_{kl}\tr[(\Pi^{[1]})^2]\,,\  \d_{ij}\d_{kl} (\tr[\Pi^{[1]}])^2\,,\  \d_{ij} \Pi^{[2]}_{kl}\,,\  \Pi^{[1]}_{ij} \Pi^{[1]}_{kl}\,,\  \d_{ij} \Pi^{[1]}_{km} \Pi^{[1]\,m}_l\,,\vs
&& \ \d_{ij} \Pi_{kl} \tr[\Pi^{[1]}]\,.
\eea
In the expansion of \refeq{Pkexp}, the ${\rm 1^{st}\ order}$ terms above are to be contracted with $\hat k^i\hat k^j$, while the ${\rm 2^{nd}\ order}$ terms are contracted with $\hat k^i\hat k^j\hat k^k\hat k^l$. 
For clarity of presentation, we have left the ${\rm 2^{nd}\ order}$ terms written above unsymmetrized. Further, note that the term $\d_{ij}\d_{kl} \tr[\Pi^{[2]}]$ is absent following the discussion above. Although we only list the corresponding terms up to second order, the expansion in \refeqs{EulBasis}{EulBasis2} can be continued to any desired order (accompanied by a proliferation of the number of terms: at third order for example, there are 14 terms). 

A very similar expansion can be performed in Lagrangian space, using the
Lagrangian deformation tensor
\be
M_{ij}(\vq) \equiv \partial_{q_i} s_j(\vq)\,.
\ee
In this case, since convective time derivatives reduce to ordinary time
derivatives in Lagrangian space, the basis can be simply constructed
out of the $m$-th order contributions $M_{ij}^{(m)}$ to $M_{ij}$. This is in contrast to the $\Pi^{[m]}_{ij}$ introduced in the Eulerian expansion above: due to the fact that convective time derivatives are nontrivial in Eulerian coordinates, the tensor $\Pi^{[m]}_{ij}$ is \emph{not} simply the $m$-th order contribution to $\Pi^{[1]}_{ij}$. As we have noted already above, the choice of Lagrangian vs. Eulerian expansions amounts to a basis choice in the same vector space, so that the final result for $\R_n$ is independent of the choice of expansion. We will used the Eulerian expansion \refeqs{EulBasis}{EulBasis2} in the following, as they are more directly related to standard perturbation theory results.

\subsection{First-order expansion}
\label{sec:exp1}

Let us begin by deriving the decomposition of the linear response $\R_1$.  From \refeq{EulBasis}, we see that there are only two relevant operators, which we parametrize as $O_{ij} = \{ \d\,\d_{ij},\,K_{ij}\}$.  Equation~(\ref{eq:Pkexp2}) then becomes
\be\label{eq:R1def}
\frac{P_m(\vk| \vx)}{P_m(k)}-1 = R_1(k) \d(\vx) + R_K(k) \khat^i\khat^j K_{ij}(\vx)\,,
\ee
where $R_1 \equiv R_\delta$. Here and in the following, any long-wavelength operator is evaluated at linear order in perturbation theory, unless explicitly stated otherwise, i.e., we drop the superscript $^{(1)}$ in the following. 
Performing a Fourier transform in the long-wavelength mode, and using that
in Fourier space,
\be
K_{ij}(\vp) = \big[p^ip^j/p^2 - \delta_{ij}/3\big]\delta(\vp)\,,
\ee
we obtain the following decomposition of $\R_1$:
\be\label{eq:R1expexp}
\R_1  \equiv  \frac1{P_m(k)} \left.\frac{{\rm d} P_m(\vk| \d(\vp))}{{\rm d} \d(\vp)}\right|_{\d(\vp)=0} = R_1(k) + R_K(k) \left(\mu^2 - \frac13\right) \,,
\ee
where $\mu = \vp\cdot\vk/(p k)$. The coefficients $R_1(k)$ and $R_K(k)$ can be derived at tree level by plugging the above definition of $\R_1$ into Eq.~(\ref{eq:Bsq}). We shall do this explicitly in the next section.

\subsection{Second-order expansion}
\label{sec:exp2}

Including all second-order operators from \refeq{EulBasis2} in \refeq{Pkexp2} (slightly rewritten to single out the density and tidal fields) and performing the Fourier transform on the two long-wavelength modes, we obtain
\bq\label{eq:PR2}
\frac{P_m\left(\vk| \d(\vp_1) \d(\vp_2) \right)}{P_m(k)}-1 &=& R_1(k)\bigg[ \d^{(2)}(\vp_1,\vp_2)\bigg] + R_K(k) \bigg[\khat^i\khat^j K_{ij}^{(2)}(\vp_1,\vp_2)\bigg]  \\
&+& \frac12 R_2(k) \bigg[\d(\vp_1)\d(\vp_2)\bigg] + R_{K \d}(k) \bigg[\khat^i \khat^j K_{ij}(\vp_1) \d(\vp_2)\bigg] \nonumber \\
&+& R_{K^2}(k)\bigg[ K_{ij}(\vp_1) K^{ij}(\vp_2)\bigg] + R_{K.K}(k) \bigg[\khat^i \khat^j K_{il}(\vp_1) K^l_{\  j}(\vp_2) \bigg] \nonumber \\ 
&+& R_{KK}(k) \bigg[\khat^i \khat^j \khat^l \khat^m K_{ij}(\vp_1) K_{lm}(\vp_2)\bigg] + R_{\Ote}(k) \bigg[\khat^i \khat^j \Ote_{ij}(\vp_1,\vp_2)\bigg]\,,
\nonumber
\eq
where $R_2 \equiv 2 R_{\d^2}$, and
\be
\Ote_{ij} \equiv \left(\frac{\partial_i\partial_j}{\lapl} - \frac13 \d_{ij}\right)\left(\d^2 - \frac32 (K_{ij})^2 \right)
\ee
is directly related to the trace-free part of $\Pi^{[2]}_{ij}$ (see Appendix C of Ref.~\cite{biasreview}). Note that we only need to consider the trace-free part of $\Pi^{[2]}$, as the trace is equivalent to other terms already taken into account.

As noted in the discussion above, the first two terms on the right-hand side of \refeq{PR2} are required by the physical definition of the first-order response, since two long-wavelength modes which combine to a second-order density field and tidal field are to be multiplied by the corresponding first-order response coefficients. This is why the above second-order expansion contains eight terms, as opposed to only six as one could naively expect from Eq.~(\ref{eq:EulBasis2}). This is analogous to the term $b_1 \d^{(2)}$ appearing in the expansion of biased tracers at second order (see Sec.~2.2 in Ref.~\cite{biasreview}). The explicit kinematic dependence encoded in \refeq{PR2} follows from standard perturbation theory, using 
\ba
\d^{(2)}(\vp_1,\vp_2) \equiv\:& F_2(\vp_1,\vp_2) \d(\vp_1) \d(\vp_2) 
= \left[\frac57 + \frac27 \mu_{12}^2 + \frac12 \mu_{12} \left(\frac{p_1}{p_2} + \frac{p_2}{p_1}\right)\right] \d(\vp_1) \d(\vp_2),
\label{eq:delta2}\\
\khat^i \khat^j K_{ij}^{(2)}(\vp_1,\vp_2) \equiv\:& \mu_1 \mu_2 \mu_{12} - \frac13 {\mu_{12}^2 } + {\frac57} \left(\frac{(\hat{\vk}\cdot\vp_{12})^2}{p_{12}^2} - \frac13 \right) (1 - \mu_{12}^2) \nonumber \\
& + \frac12  \mu_{12} \left[ (\mu_1^2 - \frac13) \frac{p_1}{p_2} + (\mu_2^2 - \frac13) \frac{p_2}{p_1}\right], \label{eq:Kij2}\\
\khat^i \khat^j \Ote_{ij}(\vp_1,\vp_2) \equiv\:& \left(\frac{(\hat{\vk}\cdot\vp_{12})^2}{p_{12}^2}-\frac13  \right) \left[ \d(\vp_1) \d(\vp_2) - \frac32 K_{lm}(\vp_1) K^{lm}(\vp_2) \right]
\ ,\label{eq:Otd}
\ea
where $\mu_1= \hat{\vk}\cdot\hat{\vp}_1$, $\mu_2= \hat{\vk}\cdot\hat{\vp}_2$ and $\mu_{12}= \hat{\vp}_1\cdot\hat{\vp}_2$.  
The kinematic shape of $\Ote_{ij}$ and $K^{(2)}_{ij}$ contains the quantity $(\hat{\vk}\cdot\vp_{12})^2/p_{12}^2$, which is a consequence of the fact that these terms are nonlocally related to $\partial_i\partial_j\Phi$ in real space. Defining $f_{12} = p_2/p_1$, the said term can be written as
\ba
\left(\frac{(\hat{\vk}\cdot\vp_{12})^2}{p_{12}^2} - \frac13 \right) (1 - \mu_{12}^2) 
&=\:
\left(\frac{( \mu_1 + f_{12} \mu_2)^2}{1 + f_{12}^2 + 2 f_{12} \mu_{12}} - \frac13 \right) (1 - \mu_{12}^2) \, \vs
&\stackrel{f_{12}=1}{=}\:
 \frac12 (\mu_1 + \mu_2)^2 (1 - \mu_{12})  - \frac13 (1 - \mu_{12}^2) \,,
\ea
where in the second equality we have specialized to the case $p_1 = p_2$. Interestingly, the rational function in the angles reduces to a polynomial shape in this particular limit. Combining Eqs.~(\ref{eq:delta2}), (\ref{eq:Kij2}), (\ref{eq:Otd}), (\ref{eq:PR2}) and (\ref{eq:respderi}), we can then write
\bq\label{eq:PR2_angle}
&&\R_2(k; \mu_1,\mu_2,\mu_{12}, f_{12}) =  R_1(k)  \Bigg[{\frac{5}{7}} +  \frac{\mu_{12}}{2}\big(f_{12} + \frac{1}{f_{12}}\big)+ \frac27 \mu_{12}^2 \Bigg] \nonumber \\
&&+ R_K(k)\Bigg[\mu_1 \mu_2 \mu_{12} - \frac13 {\mu_{12}^2 } + \frac57 \left(\frac{( \mu_1 + f_{12} \mu_2)^2}{1 + f_{12}^2 + 2 f_{12} \mu_{12}} - \frac13 \right) (1 - \mu_{12}^2)  \nonumber \\ 
&& \qquad\qquad + \frac12  \mu_{12} \Bigg( \left(\mu_1^2 - \frac13\right)f_{12} + \left(\mu_2^2 - \frac13\right)\frac{1}{f_{12}} \Bigg) \Bigg] \nonumber \\
&& + \frac12 R_2(k) + \frac12 R_{K \d}(k) \Bigg[\mu_1^2 + \mu_2^2 - \frac23 \Bigg] + R_{K^2}(k) \Bigg[\mu_{12}^2 - \frac13\Bigg]  \nonumber \\
&& + R_{K.K}(k) \Bigg[\mu_1 \mu_2 \mu_{12} {-\frac13\mu_1^2-\frac13\mu_2^2+\frac19}\Bigg] + R_{KK}(k) \Bigg[\mu_1^2\mu_2^2 - \frac13\left(\mu_1^2 + \mu_2^2\right) + \frac19\Bigg] \nonumber \\ 
&& +  \frac32 R_{\Ote}(k) \left(\frac{( \mu_1 + f_{12} \mu_2)^2}{1 + f_{12}^2 + 2 f_{12} \mu_{12}} - \frac13 \right) (1 - \mu_{12}^2)\,.
\eq
We stress again that this expression for the full second-order response $\R_2$ is valid for fully nonlinear $k$.

A similar expression has been derived in Ref.~\cite{bertolini1} using the squeezed limit of the tree-level four-point function. In particular, when restricting to $f_{12}=1$, \refeq{PR2_angle} matches Eq.~(3.16) of Ref.~\cite{bertolini1}, but with a different kinematic basis. That is, their coefficients $A_n$ are linear combinations of our $R_O$. Note however that the kinematic basis in \refeq{PR2_angle} is derived from operators that correspond to physical, local observables. In Sec.~\ref{sec:extrapolation}, we shall see that this plays an important role in guiding the extrapolation of the response coefficients to the nonlinear regime.

\section{Tree-level response coefficients}\label{sec:derivation}

In this section, we put the results of the last two sections together to explicitly derive the shape of the $R_O(k)$ response coefficients in the first-order and second-order response, by matching to the tree-level bispectrum and trispectrum, respectively.

\subsection{First order}

The first-order response coefficients $R_1(k)$ and $R_K(k)$ can be read off from the equation that is obtained by plugging \refeq{R1expexp} into \refeq{Bsq} at tree level:
\bq\label{eq:matchR1}
R_1^{\rm tree}(k) + R_K^{\rm tree}(k) \bigg(\mu^2 - \frac13\bigg) &=& 2 \lim_{p \to 0} \bigg[\fii(\vk, \vp)P_L(k) + \fii(-\vk-\vp, \vp)P_L(|\vk+\vp|) \bigg], \nonumber \\
&=& \frac{13}{7} + \left(\frac87-\fkone\right)\mu^2\,,
\eq
where we have expanded the power spectrum as $\Plin(|\vk + \vp|) \approx \Plin(k)\big(1 + \mu p \Plin^\prime(k)/\Plin(k)\big)$, and here and throughout, a prime on a power spectrum denotes a derivative with respect to $k$. Note that the terms $\propto k/p$ in $\fii$ that diverge in the limit $p\to 0$ cancel exactly. This is expected, since these terms correspond to an overall displacement due to the long mode, which is not locally observable and hence does not physically modulate the local power spectrum. The functions $R_1^{\rm tree}(k)$ and $R_K^{\rm tree}(k)$ are obtained straightforwardly by matching to the constant and the $\mu^2$ terms in Eq.~(\ref{eq:matchR1}):
\bq\label{eq:R1RK}
R_1^{\rm tree}(k) &=& \frac{47}{21} - \frac{1}{3}\fkone, \nonumber \\
R_K^{\rm tree}(k) &=& \frac{8}{7} - \fkone.
\eq
In this section, the superscript $^{\rm tree}$ serves to emphasize that the result is valid only at tree level. These expressions match the equivalent ones derived in Refs.~\cite{response} (in the case of $R_1$) and Refs.~\cite{bertolini1, 2016arXiv161104723A}. Note also that, as expected, $R_1(k) = \int_{-1}^{1} \R_1(k, \mu) {\rm d}\mu$, and that this tree-level result for $R_1$ is recovered by the simulation measurements of Ref.~\cite{response} on large scales. 

\subsection{Second order}

The derivation of the second-order response coefficients follows the exact same steps, but applied to the $n=2$ case of the tree-level limit of Eq.~(\ref{eq:Rndef}), which describes the 4-point function or trispectrum. Explicitly, we plug the $\R_2$ expansion of Eq.~(\ref{eq:PR2_angle}) into the following equation:
\bq\label{eq:sqTtree}
& \lim_{\{p_a\} \to 0} \left(\raisebox{-0.0cm}{\includegraphicsbox[scale=0.8]{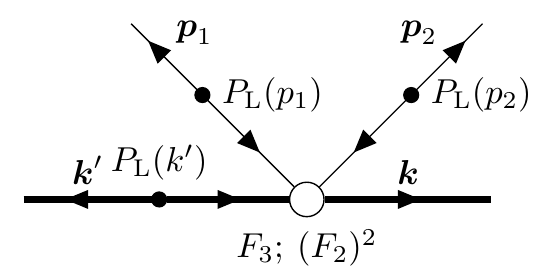}} + (\text{perm.}) \right) = \< \d(\vk) \d(\vk') \d(\vp_1) \d(\vp_2) \>_{c,\R_2}^\text{tree} \nonumber \\
& \hspace*{1cm} = 2\R_2^\text{tree}(k; \mu_{\vk,\vp_1},\mu_{\vk,\vp_2},\mu_{\vp_1,\vp_2}, p_1/p_2) \Plin(k') \Plin(p_1) \Plin(p_2) \: (2\pi)^3 \d_D(\vk+\vk'+\vp_{12})\,, \nonumber \\
\eq
where (see Appendix \ref{app:tritree}, and Appendix A.~2.~of Ref.~\cite{response} for the angle-averaged case)
\bq\label{eq:r2pk_angle1}
&& \< \d(\vk) \d(\vk') \d(\vp_1) \d(\vp_2) \>^{\text{tree}\,\prime}_{c,\R_2} = \\ 
&& \lim_{\{p_a\} \to 0} \Bigg[6\fiii(\vk, \vp_1, \vp_2) \Plin(k) \Plin(p_1)\Plin(p_2) \nonumber \\ 
&&\ \ \ \ \ \ \ \ \ \ \ \ \ \ \ \ \ \ \ \ \ \ \ \ \ \ + 4\fii(-\vp_1, \vk+\vp_1)\fii(\vp_2, \vk+\vp_1)\Plin(k + p_{1}) \Plin(p_1)\Plin(p_2)  + (\vk \leftrightarrow \vk')\Bigg]\,. \nonumber
\eq
Diagrammatically, the tree-level $\R_2^{\rm tree}$ interaction vertex can be represented as the following sum of $F_3$ and $(F_2)^2$ kernels (clarifying the meaning of the open circle in Eq.~(\ref{eq:sqTtree})):
\ba\label{eq:R2treediag}
\raisebox{-0.0cm}{\includegraphicsbox[scale=0.8]{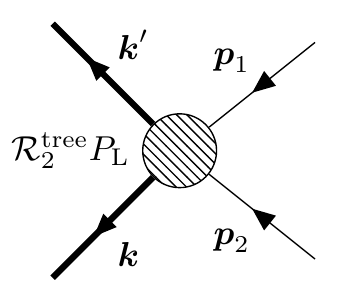}}
=\:&
\raisebox{-0.0cm}{\includegraphicsbox[scale=0.8]{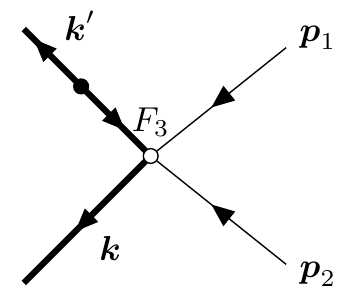}}
+
\raisebox{-0.0cm}{\includegraphicsbox[scale=0.8]{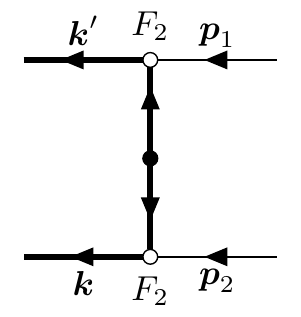}}
+ (\vk \leftrightarrow \vk')\,,
\ea
where $p_1$,$p_2$ are to be understood as much softer than $k$,$k'$. Equation (\ref{eq:r2pk_angle1}) is obtained from Eq.~(\ref{eq:R2treediag}) by the Feynman rules, after attaching propagators to the external momenta $\vp_1$, $\vp_2$.

After Taylor-expanding $\Plin(|\vk + \vp|)$ and $\Plin(|\vk + \vp_{12}|)$ in \refeq{r2pk_angle1} to second order in the amplitude of the soft modes, so that the IR divergences cancel out (cf.~Eqs.~(\ref{eq:pkexp2}) in Appendix \ref{app:tritree}), and specializing to the case $f_{12}=1$, we obtain 
\bq\label{eq:r2pk_angle}
&& \< \d(\vk) \d(\vk') \d(\vp_1) \d(\vp_2) \>^{\text{tree}\,\prime}_{c,\R_2} \approx \nonumber \\ 
&& \Bigg[ \frac{628}{147} + \frac{26}{7}\mu_{12} + \frac{16}{21}\mu_{12}^2+ \bigg(\frac{352}{147} - \frac{29}{14}\fkone\bigg)\bigg(\mu_1^2 + \mu_2^2\bigg)+  \bigg(\frac{8}{21} - \frac{3}{7}\fkone\bigg)\mu_1\mu_2  \nonumber \\
&& \qquad - \bigg(\frac{16}{7} - \frac{10}{7}\fkone\bigg)\mu_1\mu_2\mu_{12} + \bigg(\frac{656}{147} - \frac{23}{7}\fkone + \fktwo\bigg)\mu_1^2\mu_2^2 \nonumber \\
&& \qquad + \bigg(\frac{20}{21} - \frac{11}{14}\fkone\bigg)\mu_{12}\bigg(\mu_1^2 + \mu_2^2\bigg)\Bigg] \Plin(k)\Plin(p_1)\Plin(p_2)\,.
\eq
The general result for $f_{12}\neq 1$ is much more lengthy and can be found in Eq.~(\ref{eq:r2pk_angleapp}); however, the case $f_{12}=1$ is sufficient to unambigously determine all second-order response coefficients.

The second-order response coefficients can now be derived by equating Eq.~(\ref{eq:PR2_angle}), specializing to $f_{12}=1$, to Eq.~(\ref{eq:r2pk_angle}) through Eq.~(\ref{eq:sqTtree}). After matching the relevant kinematic shapes (analogously to the first-order responses above; performing this matching for a general $f_{12}$ leads to the same result), we find
\bq\label{eq:rs}
R^{\rm tree}_1(k) &=& \frac{47}{21} - \frac{1}{3}\fkone\ \ \ \ \ \ \ \ \ \ \ \ \ \ \ \ \ \ \ \ \  ; \ \ \ R^{\rm tree}_K(k) = \frac{8}{7} - \fkone; \nonumber \\
R^{\rm tree}_2(k) &=& \frac{74}{27} - \frac{22}{21}\fkone + \frac19\fktwo\ \ ; \ \ R^{\rm tree}_{K\delta}(k) = \frac{1012}{441} - \frac{41}{21}\fkone + \frac13\fktwo; \nonumber \\
R^{\rm tree}_{K^2}(k) &=& \frac{26}{441} - \frac{1}{6}\fkone\ \ \ \ \ \ \ \ \ \ \ \ \ \ \ \ \ \ \ \  ; \ R^{\rm tree}_{K.K}(k) = -\frac{44}{21} +\frac32 \fkone; \nonumber \\
R^{\rm tree}_{KK}(k) &=& \frac{328}{147} - \frac{23}{14}\fkone + \frac12\fktwo\ ; \ \ \ \ R^{\rm tree}_{\Ote}(k) = -\frac{184}{441} +\frac13\fkone\,.
\eq
As expected, we recover the first-order response coefficients via the distinct shapes of the second-order density and tidal fields, which serves as a cross-check of the calculation. Recall that, as discussed in \refsec{exp2}, the presence of these first-order coefficients in the second-order expansion is required by the fact that the response coefficients are independent of the perturbative order used to describe the long-wavelength modes.   

\section{Response coefficients in the nonlinear regime}\label{sec:extrapolation}

The response coefficients $R_O(k)$ are physical quantities that can be measured with appropriately setup N-body simulations, thereby offering a means to accurately determine them at any $k$. In the case of the isotropic response coefficients $R_n(k) = n! R_{\d^n}(k)$, which correspond to the response with respect to $\d^n$, 
the task can be carried out effectively using the separate universe simulation formalism \cite{2011ApJS..194...46G, 2011JCAP...10..031B, 2016JCAP...09..007B, wagner/etal:2014}, in which the effect of the long-wavelength density perturbation is included by following structure formation in a spatially curved Friedmann-Robertson-Walker universe. 
The latter can in turn be neatly incorporated into N-body simulations by an appropriate modification of the cosmological parameters.  The coefficients $R_n$ can be seen as the matter power spectrum analogue of the so-called  ``local-in-matter-density'' (LIMD; previously known simply as ``local'') bias parameters of large-scale structure tracers, which can similarly be measured through separate universe simulations \cite{lazeyras/etal,baldauf/etal:2015,li/hu/takada:2016}. In Ref.~\cite{response}, the authors used separate universe simulations to measure the first three isotropic response coefficients $R_1(k)$, $R_2(k)$ and $R_3(k)$ in the nonlinear regime.

The remaining response coefficients, one at first order and five at second order, which disappear under a full angle average but are relevant in general, cannot be measured using the standard separate-universe setup. Instead, simulations including large-scale tidal fields, which break isotropy, are required. Setting up such simulations, while clearly well motivated, is beyond the scope of this paper. Instead, we shall use the tree-level expressions of the previous section together with the measured isotropic response coefficients from Ref.~\cite{response} to obtain physically motivated extrapolations of all $R_O(k)$, as described in this section.

In the separate universe picture, which is exact for \lcdm\ \cite{CFCpaper2}, the isotropic responses studied in Ref.~\cite{response} can be broken down into three physically distinct contributions \cite{sherwin/zaldarriaga,r4,2014PhRvD..90j3530L,response,r10}
\begin{enumerate}
\item The {\it reference density} contribution, which accounts for the fact that the local matter power spectrum is defined with respect to the local mean density, while the global power spectrum is defined with respect to the fiducial background. This in practice corresponds to an overall rescaling of the amplitude of the power spectrum, and is uniquely determined at all orders by the perturbative evolution of the large-scale density field.
\item The {\it dilation} effect contribution, which arises because the long-wavelength modes perturb the local scale factor, which therefore rescales the size of the small-scale modes. This corresponds to a rescaling of the argument of the matter power spectrum. Note that if the long-wavelength perturbations are anisotropic, then the perturbations to the scale factor will also be anisotropic. This contribution is completely determined by the perturbative evolution of the large-scale modes, as well as the shape of the nonlinear matter power spectrum.
\item The {\it growth effect}, which quantifies the changes in structure formation induced by the long-wavelength perturbations, and corresponds to the actual physical coupling between the long- and short-wavelength modes. Reference \cite{response} defined the growth-only response coefficients $G_n(k)$, which measure the impact of this effect alone.
\end{enumerate}

Effects 1 and 2 above can be calculated exactly at any order given the nonlinear matter power spectrum in the fiducial cosmology, i.e., no dedicated separate universe simulations are necessary. Separate universe simulations are however required to measure effect 3 on nonlinear scales. Next, we will attempt to single out the contribution from the growth effect in all coefficients $R_O(k)$, and extrapolate this contribution to the nonlinear regime of structure formation using the measured growth-only isotropic responses in the separate universe simulations of Ref.~\cite{response}.

\subsection{Extrapolating the first order response coefficients}

\begin{figure}
	\centering
	\includegraphics[width=\textwidth]{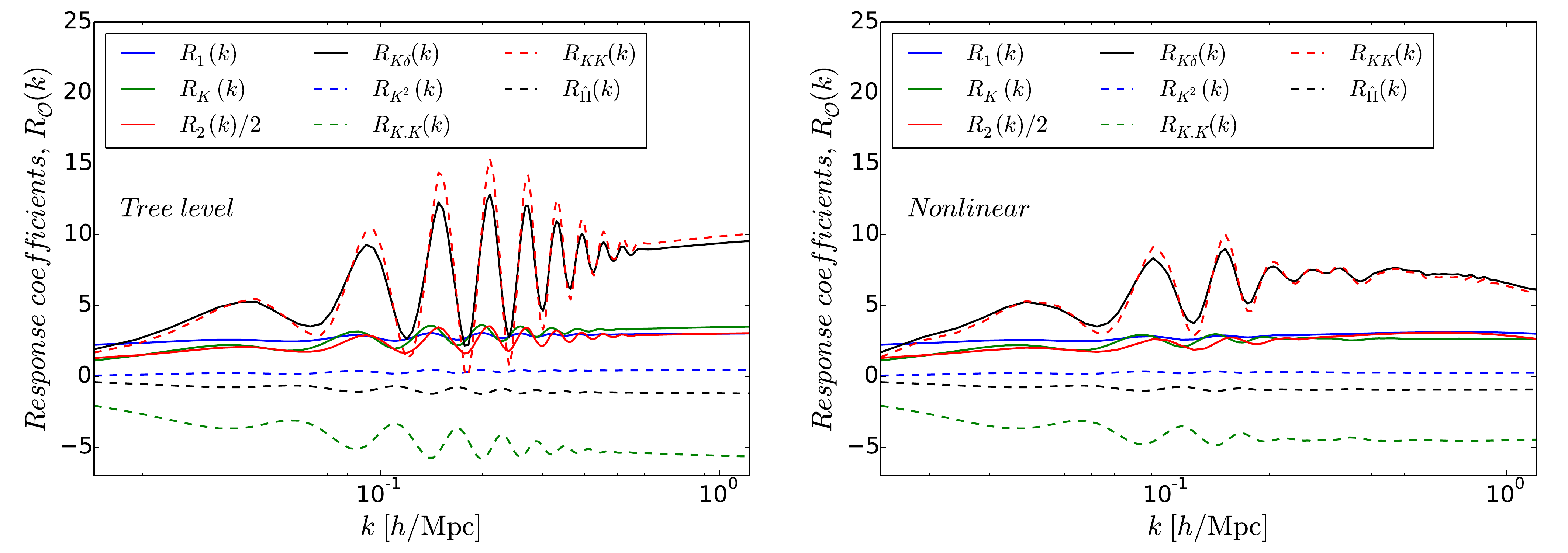}
	\caption{The left and right panels show the response coefficients $R_O(k)$ at tree level (cf.~Eqs.~(\ref{eq:rs})) and their nonlinear expression (for $R_1,\, R_2$) or extrapolation (cf.~Eqs.~(\ref{eq:rsnl})), as labeled. The results correspond to $z=0$.}
\label{fig:rs}
\end{figure}

We start with the first order response coefficients, $R_1$ and $R_K$. Following Ref.~\cite{response}, at tree level, $R_1$ can be decomposed as 
\ba
R^{\rm tree}_1(k) 
=\:& \frac{47}{21} - \frac13 \frac{d\ln \Plin(k)}{d\ln k} \vs
=\:& 1 - \frac13 \frac{d\ln \Plin(k)}{d\ln k} + \frac{26}{21} 
= 1 - \frac13 \frac{d\ln \Plin(k)}{d\ln k} + G^{\rm tree}_1(k)
\,,
\ea
where the first, second and third terms on the second line are the reference density, dilation and growth effects, respectively. At tree level, $G^{\rm tree}_1 = 26/21$. By replacing $G_1$ in this expression with its full nonlinear shape, as well as replacing the linear matter power spectrum with the nonlinear one, we obtain an expression for $R_1(k)$ which, because of the exact validity of the separate universe, perfectly matches that from the separate universe simulations \cite{response}. Here and below, we use the superscript $^{\rm tree}$ to indicate expressions valid at tree level only, and implicitly assume the general nonlinear result otherwise. The same steps can be followed to extrapolate $R_K$ to nonlinear scales as well. In particular, the tidal response at tree level is 
\be
R_K^{\rm tree}(k) = \frac87 - \frac{d\ln \Plin(k)}{d\ln k}\,.
\ee
There is no reference density effect here, and the first term corresponds to the physical growth effect of a long-wavelength tidal field \cite{tidalpaper}, while the second term is the dilation effect. Thus, in order to obtain the nonlinear $R_K$, we can adopt the following form: 
\be
R_K(k) = \alpha G_1(k) - \fkonenl\,.
\label{eq:RKnl1}
\ee
The coefficient $\alpha = 12/13$ is determined by the requirement that $R_K$ recovers the tree-level limit as $k\to 0$, so that $\alpha G_1^{\rm tree} = 8/7$. Note also that we have replaced $\Plin$ by $\Pnl$. Hence, our extrapolation for $R_K(k)$ becomes
\be
R_K(k) = \frac{12}{13} G_1(k) - \fkonenl.
\label{eq:RKnl2}
\ee
This assumes that the \emph{scale dependence} of the physical modulation, i.e. the growth effect, of the local
power spectrum by a long-wavelength tidal field is the same as that of a
long-wavelength density perturbation. Beyond the large-scale limit, where
the tree-level result is exact, this expression is not expected to be perfect. 
In the halo model description of the nonlinear matter density (see Ref.~\cite{cooray/sheth} for a review), the scale dependence of the growth response is determined by the sensitivity of the abundance of massive halos to long-wavelength perturbations, which leads to the increased response on scales of $k \sim 0.7\iMpch$ (see e.g.~Fig~2 of Ref.~\cite{response}). On the other hand, the structure of halos, which determines the power spectrum in the deeply nonlinear ``1-halo'' regime, is largely insensitive to long-wavelength perturbations, which suppresses the response at $k \gtrsim 1\iMpch$. These considerations, which are described in more detail in Ref.~\cite{response}, are expected to hold for anisotropic large-scale perturbations as well. Thus, our extrapolation should serve as a reasonable first approximation that has no free parameters. The tree-level and nonlinear shapes of $R_1(k)$ and $R_K(k)$ are shown in \reffig{rs}.

\subsection{Extrapolating the second order response coefficients}

The same considerations as above apply, but become more involved in the case of the second-order anisotropic response coefficients, $R_{K\delta}$, $R_{K^2}$, $R_{K.K}$, $R_{KK}$ and $R_{\Ote}$. We can split these coefficients into those that do not depend on the second derivative of the matter power spectrum and those that do. For those that do not, which are $R_{K^2}$, $R_{K.K}$ and $R_{\Ote}$, we choose to extrapolate them in the exact same way as $R_K$ above, i.e., by using $G_1(k)$ to rescale the constant term.\footnote{We have checked that using $G_2(k)$ instead gives the same result up to $10-20\%$ for $k \lesssim 1\kunit$. We have also checked that assuming a reference density term to extrapolate these coefficients makes practically no difference up to $k\approx 0.3\kunit$, and never induces differences larger than $20\%$ for higher values of $k$.} Those that do depend on the second derivative, which are $R_{K\delta}$ and $R_{KK}$, require more deliberation. We can use clues provided by the decomposition of $R_2$ into reference density, dilation, and growth contributions derived in Ref.~\cite{response} (using their Eulerian definitions, not Lagrangian ones, cf.~\refapp{RLag}):
\bq\label{eq:R2dec}
R_2(k) = \left(\frac{8}{21} G_1(k) + G_2(k)\right )+ \left(-\frac{2}{9} - \frac23 G_1(k)\right)\fkonenl + \frac19\fktwonl - \frac23kG_1^{\prime}(k). \nonumber \\
\eq
The term $G_1'(k)$ vanishes at tree level and only accounts for less than $\approx 10\%$ of the total value of $R_2(k)$ on small scales. Here, we ignore this term and take the following ansatz for the extrapolation of $R_{K\delta}$ and $R_{KK}$:
\bq\label{eq:2derivativeextraplation}
\label{eq:2derivativeextraplation1} R_{K\delta} &=& \beta\left[\frac{8}{21}G_1(k) + G_2(k)\right] + \gamma \left(-\frac{2}{9} - \frac23 G_1(k)\right) \fkonenl + \frac13\fktwonl, \\
\label{eq:2derivativeextraplation2} R_{KK} &=& \beta\left[\frac{8}{21}G_1(k) + G_2(k)\right] +\gamma \left(-\frac{2}{9} - \frac23 G_1(k)\right) \fkonenl + \frac12\fktwonl,
\eq 
where the coefficients $\beta$ and $\gamma$ are determined by the conditions that $\beta\left[\frac{8}{21}G_1^{\rm tree} + G_2^{\rm tree}\right]$ and $\gamma \left(-\frac{2}{9} - \frac23 G_1^{\rm tree}\right)$ match the corresponding coefficients $\propto 1$ and $\propto \Plin'(k)/\Plin(k)$ of the tree-level expressions in Eq.~(\ref{eq:rs}). Note that $\beta$ and $\gamma$ take on different values in $R_{K\delta}$ and $R_{KK}$, and $G_2^{\rm tree} = 3002/1323$. We emphasize again that there are no free parameters in the resulting extrapolated $R_O(k)$.

Putting this all together, the nonlinear extrapolations of the response coefficients are given by
\bq\label{eq:rsnl}
R_K(k) &=& \frac{12}{13} G_1(k) - \fkonenl, \nonumber \\
R_{K\delta}(k) &=& \frac{1518}{1813}\left[\frac{8}{21}G_1(k) + G_2(k)\right] + \frac{41}{22}\left[-\frac29 - \frac23G_1(k)\right]\fkonenl + \frac13\fktwonl \nonumber \\
R_{K^2}(k) &=& \frac{1}{21}G_1(k) - \frac{1}{6}\fkonenl,\nonumber \\ 
R_{K.K}(k) &=& -\frac{22}{13}G_1(k) +\frac32 \fkonenl, \nonumber \\
R_{KK}(k) &=& \frac{1476}{1813}\left[\frac{8}{21}G_1(k) + G_2(k)\right] + \frac{69}{44}\left[-\frac29 - \frac23G_1(k)\right]\fkonenl + \frac12\fktwonl \nonumber \\
R_{\Ote}(k) &=& -\frac{92}{273}G_1(k) +\frac13\fkonenl\,,
\eq
where $G_1(k)$ and $G_2(k)$, as well as $R_1(k)$ and $R_2(k)$, are taken directly from the measurements of Ref.~\cite{response}. We note that these extrapolations assume similar scale dependencies of the response coefficients for different operators, which is an approximation that should be assessed using simulations implementing the different long-wavelength configurations.

The tree-level and nonlinear extrapolation of the response coefficients are shown in Fig.~\ref{fig:rs} at $z=0$. We have used CAMB \cite{camb} to evaluate $\Plin(k)$, the Coyote emulator \cite{emulator} to evaluate $P_m(k)$, and compute the derivatives via finite-differencing. By construction, the curves on the left and right panels agree on large scales. When compared to the tree-level result, the nonlinear expressions exhibit the expected suppression of the amplitude of the BAO oscillations, and the broad-band suppression on small scales as one begins to probe the interior of virialized structures. Further, one can identify the following hierarchy between the response coefficients:
\bq
\{R_{K \d},\,R_{KK}\} > \{R_{2}/2, R_K, R_1, -R_{K.K}\} > \{R_{K^2}, -R_{\Ote}\}.
\eq
This hierarchy is determined by the tree-level result, and it is preserved by our nonlinear extrapolation. This again should also be confirmed by full simulation measurements of the anisotropic response coefficients.

\section{Application: squeezed matter power spectrum covariance}\label{sec:sqcov}

In this section, we apply the response formalism described in the previous sections to predict the non-Gaussian part of the matter power spectrum covariance in the squeezed limit. We focus on the frequently considered case of the equal-time power spectrum covariance here, although this can be straightforwardly generalized to the covariance of power spectra at different times (relevant for line-of-sight integrated observables such as lensing). The power spectrum covariance is defined by
\bq\label{eq:covdef}
\cov(\vk_1,\vk_2) \equiv \big< \hat{P}_m(\vk_1) \hat{P}_m(\vk_2) \big> - \big< \hat{P}_m(\vk_1) \big>\big< \hat{P}_m(\vk_2) \big>
\eq
where the hat on top of the power spectra indicates that these are power spectrum estimators. Specifically, the estimators average over a set of Fourier modes within bins centered on $\vk_1,\,\vk_2$ (we use the central wavenumbers $\vk_1,\vk_2$ to label the bins, for clarity).  The covariance can be decomposed into three contributions:
\bq
\cov(\vk_1,\vk_2)
= \cov^{\rm G}(\vk_1,\vk_2) + \cov^{\rm NG}(\vk_1,\vk_2) + \cov^{\rm SS}(\vk_1,\vk_2)\,,
\label{eq:covcomp}
\eq
described in the following.
\begin{enumerate}
\item The Gaussian contribution $\cov^{\rm G}$. This represents the trivial disconnected diagonal contribution to \refeq{covdef}. If we consider the covariance of the angle-averaged power spectrum within a wavenumber bin of width $\Delta k$, the Gaussian contribution is given by
\be
\cov^{\rm G}(k_1,k_2) = \frac{2}{N_k} P_m(k_1)P_m(k_2)\delta_{k_1k_2}\,,
\ee
where $N_k = V V_k/(2\pi)^3$ is the number of modes that are averaged over, with $V_k = 4\pi k^2\Delta k$ and $V$ is the survey volume.
\item The non-Gaussian (trispectrum) contribution, $\cov^{\rm NG}$, which (assuming Gaussian initial conditions) is induced by mode-coupling due to nonlinear gravitational evolution. Specifically, this involves the ``parallelogram'' configuration of the trispectrum, $T(\vk_1, -\vk_1, \vk_2, -\vk_2)$, and will be described in more detail below. 
\item The super-sample covariance contribution $\cov^{\rm SS}$ \cite{2007NJPh....9..446T, 2009ApJ...701..945S, takada/hu:2013, li/hu/takada, 2014PhRvD..90j3530L}. This second non-Gaussian contribution accounts for the interaction of modes observable within the survey with modes whose wavelength is larger than the scale of the survey \cite{takada/hu:2013}. This contribution is in fact completely captured by the first-order response $\R_1$ \cite{li/hu/takada,2016arXiv161104723A}. For the case of angle-averaged spectra and isotropic survey window functions one obtains
\be
\cov^{\rm SS}(k_1,k_2) = \Bigg[\frac{1}{V^2} \int\frac{{\rm d}^3\vp}{(2\pi)^3} |\tilde{W}(p)|^2\Plin(p)\Bigg] R_1(k_1) P_m(k_1) R_1(k_2) P_m(k_2)\,,
\ee
where $\tilde{W}(p)$ is the Fourier transform of the survey window function. This contribution formally arises from the convolution of the trispectrum with the window function, and scales differently with survey volume than the other two contributions. This is why it is sensible to treat it separately from the parallelogram trispectrum that contributes to $\cov^{\rm NG}$. Note also that the super-sample contribution is not included when estimating the power spectrum covariance from an ensemble of standard N-body simulations, which do not include fluctuations on scales larger than the simulation box.
\end{enumerate}
We note in passing that when estimating the covariance from N-body simulations, there is also a purely numerical contribution from the particle shot noise, which we have not included here.

Both the first and third contributions in \refeq{covcomp} are well understood. The second contribution, $\cov^{\rm NG}$, is however considerably more challenging to predict. For the case of angle-averaged power spectra, this contribution is given as a bin-average over the trispectrum through 
\bq\label{eq:covngdef}
\cov^{\rm NG}(k_1,k_2) &=& \frac{1}{V} \int_{V_{k_1}}\frac{{\rm d}^3\vk}{V_{k_1}} \int_{V_{k_2}}\frac{{\rm d}^3\vk'}{V_{k_2}} T(\vk, -\vk, \vk', -\vk') \approx V^{-1} \int_{-1}^{1}\frac{{\rm d}\mu_{12}}{2} T(\vk_1, -\vk_1, \vk_2, -\vk_2), \nonumber \\
\eq
where $\mu_{12} ={\vk}_1\cdot{\vk}_2/(k_1 k_2)$ and the simplifying approximation follows from assuming very narrow bins ($\Delta k/k_i \ll 1$) around $\vk_1,\,\vk_2$ such that the trispectrum can be treated as constant over the bins (see e.g.~Ref.~\cite{mohammed1} for an explicit demonstration showing that this is valid). Equation (\ref{eq:covngdef}) provides the precise relation between the covariance of the angle-averaged power spectrum and the matter trispectrum. Moreover, the trispectrum configuration appearing in \refeq{covngdef} is a function of $k_1$, $k_2$, and the angle $\mu_{12}$ between them. We can thus generalize \refeq{covngdef} and define the non-Gaussian power spectrum covariance before angle-averaging, $\Cov^{\rm NG}(k_1,k_2,\mu_{12}) = T(\vk_1, -\vk_1, \vk_2, -\vk_2)/V$ (distinguished here from $\cov^{\rm NG}(k_1,k_2)$ by the presence of $\mu_{12}$ in the argument), which we can decompose in multipoles as
\ba
\cov^{\rm NG}(k_1,k_2,\mu_{12}) =\:& \sum_{\ell=0,2,4,\cdots} \cov^{\rm NG}_{\ell}(k_1,k_2) \P_\ell(\mu_{12})\,,\quad\mbox{where} \vs
\cov^{\rm NG}_\ell(k_1,k_2) =\:& \frac{2\ell+1}{2} \int_{-1}^1 d\mu_{12}\, V^{-1}T(\vk_1, -\vk_1, \vk_2, -\vk_2) \P_\ell(\mu_{12})\,,
\label{eq:CNGell}
\ea
and $\P_\ell(x)$ is the Legendre polynomial of order $\ell$. Note that due to the fact that $P_m(-\vk)=P_m(\vk)$ has to hold, the odd multipole moments vanish identically. Note also that the $\ell = 0$ case (called the monopole) corresponds exactly to $\cov^{\rm NG}(k_1, k_2)$ in Eq.~(\ref{eq:covngdef}). This is the most commonly considered case of the power spectrum covariance, i.e., that of the angle-averaged power spectrum.

Here, we shall be interested in the squeezed limit of $\cov^{\rm NG}(k_1, k_2, \mu_{12})$, where $k_1 \ll \min\{k_2,\,\knl\}$, i.e., the non-Gaussian covariance of the small-scale with the large-scale power spectrum. Naturally, by symmetry, the same results trivially hold in the limit $k_2\ll \min\{k_1,\,\knl\}$.  Using \refeq{covngdef} as well as \refeq{sqnpt} for $n=2$, one can write
\ba
\lim_{\substack{k_1\to0 \\ {\rm or} \\k_2\to0}}\cov^{\rm NG}(k_1,k_2,\mu_{12}) = \:& V^{-1}\,2 \R_2(\khard; \mu_{12}, -\mu_{12}, -1, 1) [\Plin(\ksoft)]^2 P_m(\khard) \vs 
&+ \O\left(\frac{\ksoft^2}{\khard^2},\  \frac{\ksoft^2}{\knl^2} \right) \vs
\equiv\:& \cov^{\rm NG,sq}(k_1,k_2,\mu_{12}) 
+ \O\left(\frac{\ksoft^2}{\khard^2},\  \frac{\ksoft^2}{\knl^2} \right)
\,,
\label{eq:sqcov1}
\ea
where $\khard = \max\{k_1, k_2\}$, $\ksoft = \min\{k_1, k_2\}$. Note that the response-type contribution $\propto\R_1$ of \refeq{sqTR1} does not contribute in this particular configuration because $p_{12}=0$. The leading corrections come from beyond-squeezed-limit as well as from nonlinear terms, as discussed in \refsec{response} and \refsec{biasexp}. Some of the nonlinear terms involve loop interactions that are not captured by the $\R_2$ interaction vertex alone, and which become important if $\ksoft$ is not sufficiently smaller than $\knl$. Here, we work only at tree level in $\ksoft$, whose leading contribution is the first term in the second line of Eq.~(\ref{eq:sqcov1}). For completeness, we note also that \refeq{sqcov1} can naturally be generalized to the unequal-time covariance as
\ba\label{eq:sqcov1time}
\cov^{\rm NG,sq}(k_1,t_1;k_2,t_2;\mu_{12})  =\:& V^{-1}\,2 \R_2(\khard,t_{\rm hard}; \mu_{12}, -\mu_{12}, -1, 1) \nonumber \\
& \times [\Plin(\ksoft,t_{\rm soft})]^2 P_m(\khard,t_{\rm hard})\,,
\ea
where $t_{\rm hard}$ and $t_{\rm soft}$ correspond to the times of the hard and soft momenta, respectively. Using Eq.~(\ref{eq:PR2_angle}), we can write the configuration of $\R_2$ that enters in \refeq{sqcov1} as
\bq\label{eq:sqcov}
2\R_2(k, \mu_{12}, -\mu_{12}, -1, 1)
&=&R_2(k) + \frac{4}{3}R_{K^2}(k) + \frac{4}{9}R_{K.K}(k) \nonumber \\
&+& \Bigg[ R_{K\delta}(k) + \frac{R_{K.K}(k)}{3}\Bigg]\left(\frac{4}{3}\P_2(\mu_{12})\right) + 2R_{KK}(k) \left(\frac{2}{3}\P_2(\mu_{12})\right)^2, \nonumber \\
\eq
where $\P_2(\mu_{12})=(3\mu_{12}^2 - 1)/2$ is the second Legendre polynomial. Recall that \refeqs{sqcov1}{sqcov1time} are exact in the limit where the soft mode is sufficiently small, with any restrictions placed on the hard mode coming only from the maximum wavenumbers probed by the simulations which provide the response coefficient measurements.

Naturally, the multipole expansion of \refeq{CNGell} remains valid in the squeezed limit, and so we can define
\bq
\label{eq:covmultipoles}\cov_{\ell}^{\rm NG,sq}(k_1,k_2) &=&
V^{-1}[\Plin(\ksoft)]^2 P_m(\khard) 
\frac{2\ell+1}{2} \nonumber \\
&& \ \times \int_{-1}^{1}
2 \R_2(\khard; \mu_{12}, -\mu_{12}, -1, 1) 
\P_\ell(\mu_{12}) {\rm d}\mu_{12}\,. \nonumber \\
\eq
Given the kinematic dependences appearing in Eq.~(\ref{eq:sqcov}), only $\ell=0,2,4$ are non-vanishing. These multipoles are given as linear combinations of $R_O(k)$ as we discuss next.

\begin{figure}
	\centering
	\includegraphics[width=\textwidth]{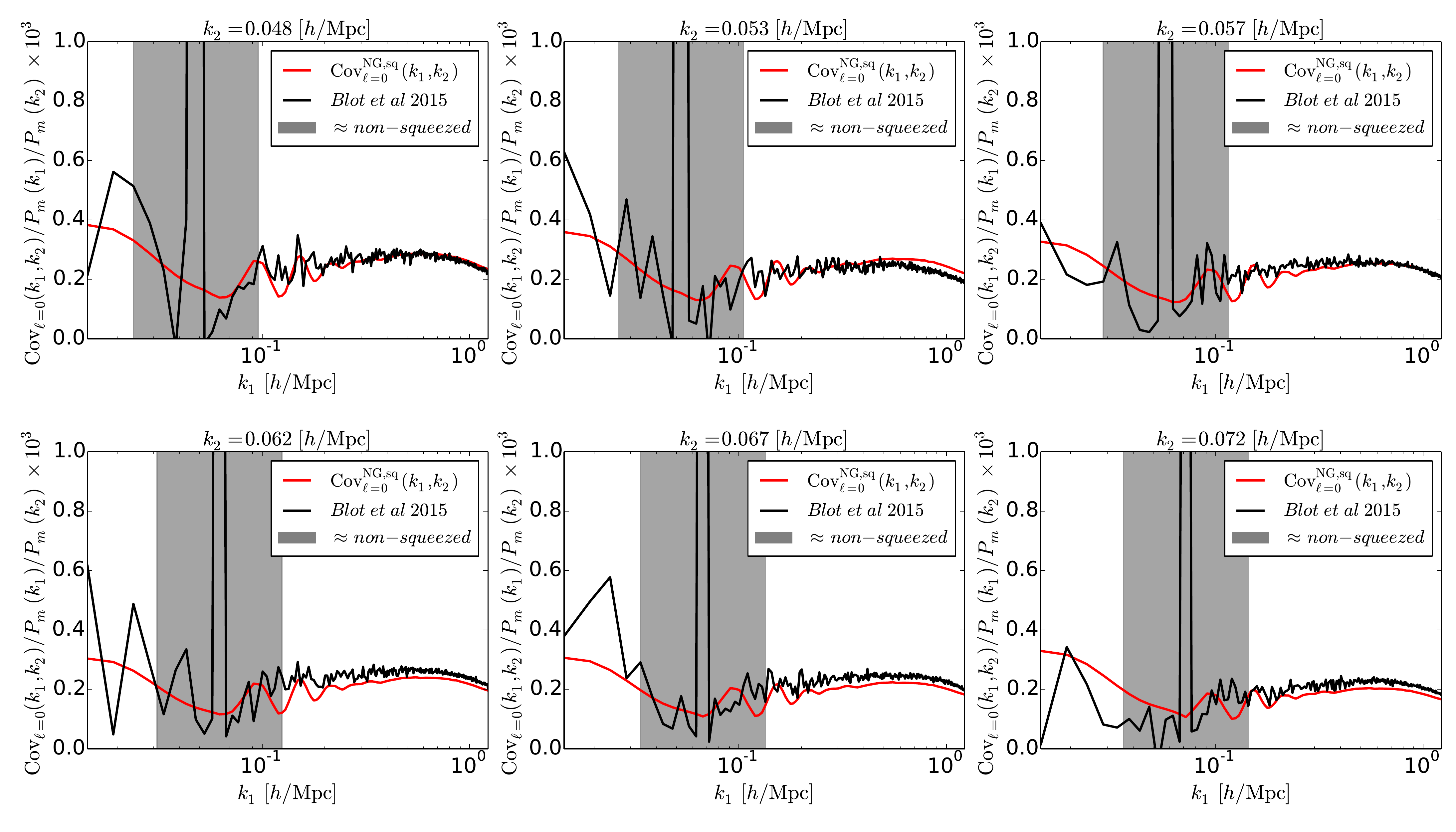}
	\caption{Angle-averaged matter power spectrum covariance $\cov_{\ell=0}(k_1,k_2)$, plotted as a function of $k_1$ for the fixed values of $k_2$ indicated above each panel. The solid black line shows the covariance matrix estimated from N-body simulations in Ref.~\cite{blot2015}. The red line shows the result of Eq.~(\ref{eq:monocov}) for the monopole squeezed-limit covariance $\cov^{\rm NG, sq}_{\ell=0}(k_1,k_2)$. The shaded area in each panel covers the region where $1/2 \leq k_1/k_2\leq 2$. The prediction shown in red is only expected to hold outside of the shaded area.}
\label{fig:sqcov}
\end{figure}

Taking $\ell=0$ in Eq.~(\ref{eq:covmultipoles}), we arrive at the following expression for the monopole squeezed-limit covariance,
\bq\label{eq:monocov}
\cov_{\ell = 0}^{\rm NG,sq}(k_1,k_2)
&=&V^{-1}\bigg[R_2(\khard) + \frac{4}{3}R_{K^2}(\khard) + \frac{4}{9}R_{K.K}(\khard) + \frac{8}{45}R_{KK}(\khard)\bigg]\nonumber \\ 
&& \times P_{m}(\khard)[\Plin(\ksoft)]^2\,.
\eq
When inserting the tree-level expressions for the $R_O(k)$ derived in \refsec{derivation} into \refeq{monocov}, we recover exactly the tree-level squeezed covariance derived in Ref.~\cite{bertolini1}.

\bigskip

In Fig.~\ref{fig:sqcov}, we show \refeq{monocov} (red line), as well as the angle-averaged covariance matrix estimated using N-body simulations in Ref.~\cite{blot2015} (black; see e.g.~Refs.\cite{2009ApJ...700..479T, joachimpen2011, 2011ApJ...734...76S, li/hu/takada} for other estimations of the covariance from simulations). The result corresponds to $z=0$ and we take $V = 656.25\:h^{-3} {\rm Mpc}^3$, which is the volume of the simulation boxes. The covariance matrix is shown as a function of $k_1$ for the fixed $k_2$ values indicated above each panel. Note that in the figure we are not restricting to squeezed configurations, i.e., we also show the result for $k_1 \approx k_2$, which is a regime in which the result is not expected to be accurate. The grey shaded area shows the region where $1/2 \leq k_1/k_2 \leq 2$, which serves to mark roughly these non-squeezed configurations. Moreover, note also that in each panel, the meaning of $\khard$ and $\ksoft$ switches on either side of the $k_1=k_2$ equality. Specifically, when $k_1<k_2$ (left of the center of the grey area) $\khard = k_2$, whereas $\khard = k_1$ when $k_1 > k_2$ (right of the center of grey area).

Figure~\ref{fig:sqcov} shows that Eq.~(\ref{eq:monocov}) provides a good description of both the amplitude and shape of the simulation results for all configurations shown. For $\ksoft = k_2 \lesssim 0.06 \iMpch$ and $\ksoft < \khard/2$ (i.e., outside of the shaded area), the agreement is remarkable. In this squeezed regime, we expect all other contributions to the total covariance to be negligible, such that comparisons to the result of Eq.~(\ref{eq:monocov}) may be a useful diagnostic of any systematics in covariance estimates based fully on N-body simulations. For higher values of $k_2$, shown in the lower panels, one can observe that our result slightly underpredicts the simulation results at high $k_1$. This is due to nonlinear corrections which become noticeable as $\ksoft/\knl$ becomes sizeable. By including the 1-loop contribution to the covariance, we expect to obtain a significantly improved match.

Note also that the response coefficients were measured in Ref.~\cite{response} for a slightly different cosmology than the one used for the covariance measurements of Ref.~\cite{blot2015}.  In particular, Ref.~\cite{response} used: $h = 0.7$, $\Omega_mh^2 = 0.1323$, $\Omega_bh^2 = 0.023$, $n_s = 0.95$, $\sigma_8(z=0) = 0.8$. We have verified that the results obtained using the cosmology of Ref.~\cite{response} to compute power spectra are indistinguishable from those using the cosmology of Ref.~\cite{blot2015}. Relative to the power spectrum, we expect the responses to be less sensitive to changes in cosmological parameters, hence we expect that this slight inconsistency in the parameters is entirely negligible for our comparisons here.

\begin{figure}
	\centering
	\includegraphics[width=\textwidth]{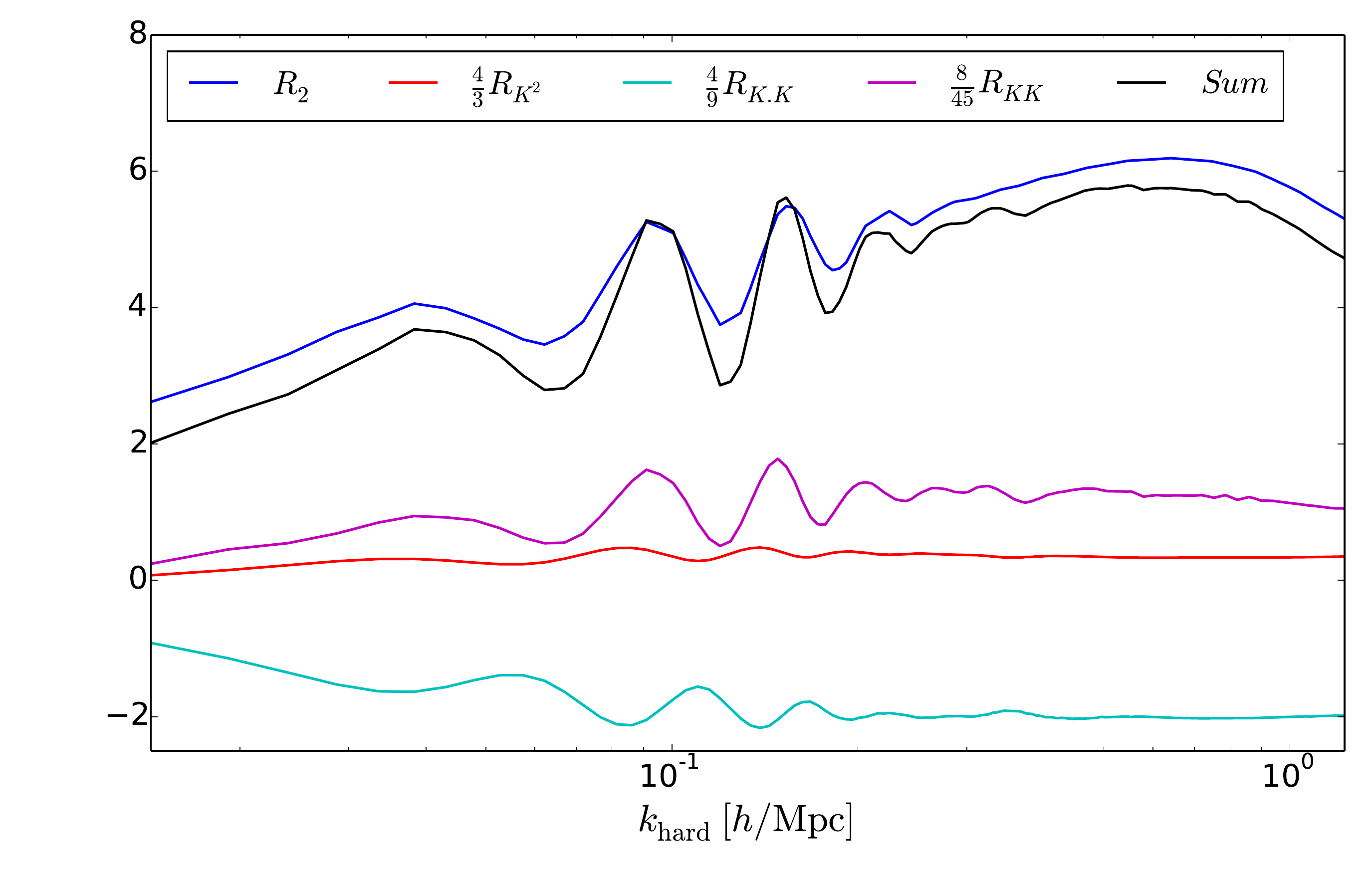}
	\caption{Contributions to the squeezed-limit monopole covariance, Eq.~(\ref{eq:monocov}), from each of the response coefficients, as labeled. The actual squeezed covariance is obtained from the black curve by multiplying with $V^{-1}P_m(\khard)[\Plin(\ksoft)]^2$.}
\label{fig:monocont}
\end{figure}

Figure \ref{fig:monocont} shows the relative magnitude of the response coefficient terms that contribute to the monopole squeezed-limit covariance in Eq.~(\ref{eq:monocov}). The figure shows that, although the isotropic second-order response coefficient $R_2$ provides the dominant contribution (blue line), the other response coefficients still contribute appreciably. This highlights the importance of the anisotropic responses, even when considering the angle-averaged covariance. Recall that we have relied on extrapolations for the anisotropic responses (cf.~Sec.~\ref{sec:extrapolation}), as these have so far not been measured directly in N-body simulations. The fact that their contribution is non-negligible, together with the good agreement between model and simulations displayed in Fig.~\ref{fig:sqcov}, then suggests that our extrapolation steps are, at the very least, not grossly wrong.

One might also wonder how well the prediction performs in \reffig{sqcov} when using the tree-level (cf.~Eqs.~(\ref{eq:rs})) rather than fully nonlinear response coefficients (cf.~Eqs.~(\ref{eq:rsnl})). We find that the lack of both BAO damping and suppression at high $k$ of the tree-level coefficients leads to a significantly worse description of the covariance, as expected.

\begin{figure}
	\centering
	\includegraphics[width=\textwidth]{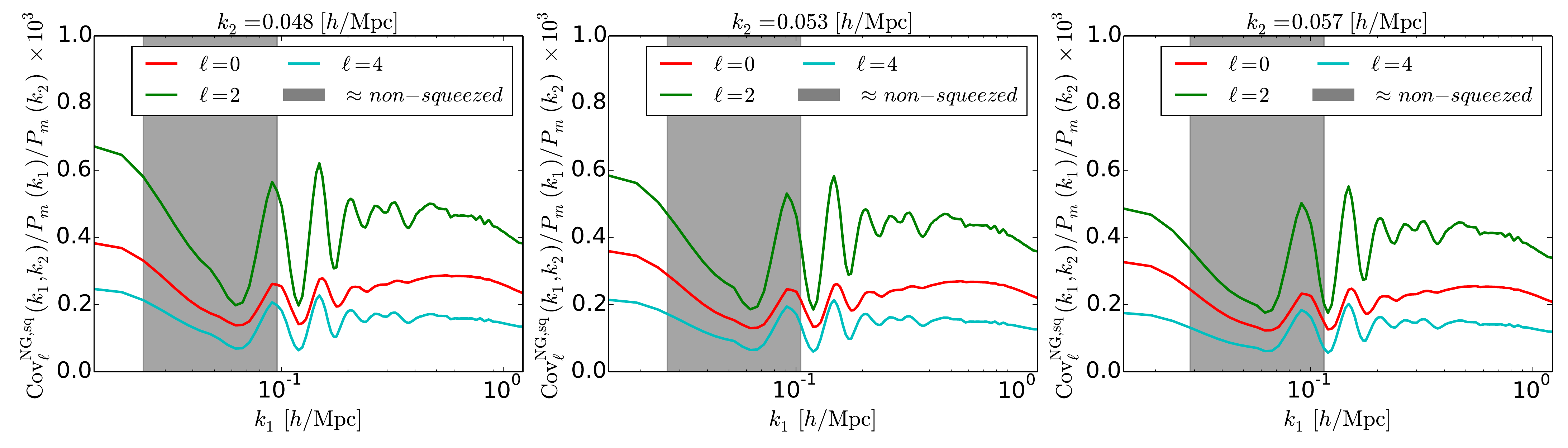}
	\caption{Same as Fig.~\ref{fig:sqcov}, but showing also the quadrupole ($\ell=2$, Eq.~(\ref{eq:quadcov})) and hexadecupole ($\ell=4$, Eq.~(\ref{eq:hexacov})). All other multipoles vanish identically.}
\label{fig:quadcov}
\end{figure}

Finally, we give the quadrupole ($\ell = 2$) and hexadecupole ($\ell = 4$) of the squeezed-limit covariance, obtained from Eq.~(\ref{eq:covmultipoles}):
\bq\label{eq:quadcov}
\cov_{\ell = 2}^{\rm NG,sq}(k_1,k_2) &=& V^{-1}\bigg[\frac43 R_{K\delta}(\khard) + \frac49 R_{K.K}(\khard) + \frac{16}{63} R_{KK}(\khard)\bigg]  \\
& & \  \times P_{m}(\khard)[\Plin(\ksoft)]^2, \nonumber \\
\label{eq:hexacov} \cov_{\ell = 4}^{\rm NG,sq}(k_1,k_2) &=& V^{-1}\bigg[\frac{16}{35} R_{KK}(\khard)\bigg] P_{m}(\khard)[\Plin(\ksoft)]^2\,.
\eq
As expected, the isotropic responses $R_n(k)$ do not contribute for $\ell > 0$. Figure \ref{fig:quadcov} compares the $l=0,2,4$ cases. We see that, for the squeezed covariance, the quadrupole is predicted to be larger than the monopole. These results reiterate the power and usefulness of the response approach, which immediately allows us to derive all multipoles of the squeezed matter power spectrum covariance. We leave the comparison of the higher multipoles of the power spectrum covariance with simulation results (e.g.~Ref.~\cite{joachimpen2011}) to future work.

\section{Summary and conclusions}\label{sec:conc}

We have presented a rigorous definition of power spectrum responses $\R_n$, which can be used 
to directly describe the squeezed limit of $(n+2)$-point functions with
two hard and $n$ soft modes. More specifically, the dominant
tree-level contributions to the $(n+2)$-point matter correlation functions
in this limit are those that are linear in the hard power spectrum (cf.~Sec.~\ref{sec:genn} and Appendix \ref{app:tritree}), and these are precisely the terms
which can be represented through power spectrum responses. The response approach can be employed whenever long-short mode
coupling is relevant in perturbation theory. The response functions can be measured accurately in the nonlinear regime with only a few N-body simulations, and as a result, by replacing the relevant interactions in perturbation theory with the corresponding response vertices (cf.~Eq.~(\ref{eq:Rndef})), one effectively {\it resums} infinitely many higher loop contributions (of a specific form), dramatically extending the range of scales where perturbative approaches can be employed and are predictive. 

The $\R_n$ can be decomposed
into a finite set of response coefficients $R_O$ (which at fixed time only depend on
the hard mode $k$) that multiply well-defined kernels that specify the kinematic dependences and are fully
determined by perturbation theory. This organization of the structure of the responses simplifies their description significantly (cf.~Sec.~\ref{sec:biasexp}).  The determination of the kernels follows from a complete enumeration of local gravitational observables, which was recently
developed in the context of general perturbative bias expansions \cite{senatore:2014,MSZ}. The shape of the $R_O$ can be determined at tree level by matching to the corresponding squeezed matter correlation functions (cf.~Sec.~\ref{sec:derivation}). On nonlinear scales, their shape can be determined with the aid of dedicated small-volume simulations, as was shown recently for
the isotropic response coefficients in Ref.~\cite{response}.

As an application of this framework, we have considered the non-Gaussian (connected) part of the matter power spectrum covariance. This is a crucial component in the cosmological interpretation of, for example, the two-point function of gravitational lensing shear \cite{2001ApJ...554...56C, 2009ApJ...701..945S, 2013MNRAS.429..344K, 2013PhRvD..87l3538S}, and several steps in modeling it have been made recently (see e.g.~Ref.~\cite{2016PhRvD..93l3505B} for a perturbation theory calculation at 1-loop order, and Refs.~\cite{mohammed/seljak,mohammed1} for a more phenomenological approach).
Here, we have considered the squeezed limit of the power spectrum covariance $\cov^{\rm NG}(\ksoft,\khard)$ at tree level in the soft mode, i.e. for $\ksoft\ll\knl$, $\ksoft\ll\khard$, but any $\khard$. This covariance corresponds to a configuration of the matter trispectrum whose relevant contributions are completely determined by $\R_2$ (cf.~Sec.~\ref{sec:sqcov}). The response formalism therefore allows us to readily evaluate the covariance of the angle-averaged power spectrum, as well as its anisotropic part. The interpretation of this squeezed limit of the covariance as a specific response had been already pointed out by Ref.~\cite{bertolini1}. The specific decomposition of $\R_2$ into a set of physical response coefficients that we performed here, however, has the advantage of permitting the use of physical considerations to estimate the nonlinear behavior of the response. This includes guiding the design of the simulation setups that are needed to measure the anisotropic response coefficients, as well as guiding the extrapolation of the anisotropic $R_O$ to nonlinear scales using the isotropic responses measured in simulations (cf.~Sec.~\ref{sec:extrapolation}).

We have compared the resulting prediction of the angle-averaged $\cov^{\rm NG}(\ksoft, \khard)$ in the squeezed limit to fully numerical estimates of the covariance, and found very good agreement when the soft mode is sufficiently linear, $\ksoft \lesssim 0.06\ \kunit$ (cf.~Sec.~\ref{sec:sqcov}). For higher values of $\ksoft$, terms beyond tree level become relevant. We stress that our predictions here have relied on an extrapolation, rather than measurement, of the anisotropic response coefficients in the nonlinear regime. Indeed, this is the only approximation made in our prediction for the squeezed-limit covariance that is relevant when comparing with gravity-only simulations. A simulation measurement of these will require nontrivial technical work, since the anisotropy is incompatible with standard cubic simulation boxes. Nevertheless, the actual measurement of the coefficients will involve substantially less computational resources compared to full numerical measurements of the covariance (e.g., compare the total simulation volume used in Ref.~\cite{response} to that in Ref.~\cite{blot2015}). Given such measurements, the prediction of the squeezed-limit covariance becomes \emph{exact}.  The response formalism applied to the covariance remains true even beyond the gravity-only case we considered so far. A measurement of the responses, and hence squeezed covariance, including baryonic effects is entirely feasible due to the greatly reduced computational demands. For the same reasons, estimating the covariance for a range of cosmological models would also be within reach. The issue of how responses can be useful for the covariance beyond the squeezed limit is subject of ongoing work.

We finish with a brief list on possible further applications and extensions of the response approach:
\begin{itemize}
\item Covariance of power spectra in redshift space, relevant for the application to spectroscopic tracers, and at different times (cf.~Eq.~(\ref{eq:sqcov1time})), relevant for applications to weak lensing measurements.
\item Response of the matter bispectrum: this can be defined in close analogy with the power spectrum response defined in \refsec{response}, and corresponds to resummed vertices with three hard outgoing lines $\vk_1,\vk_2,\vk_3$ and $n$ ingoing soft lines $\vp_a$. The main difference to the power spectrum case is that the kinematic structure becomes more complicated.
\item Response of the galaxy power spectrum: there is in principle no obstacle to applying this formalism to biased tracers instead of matter, for example to predict the galaxy power spectrum covariance \cite{2013MNRAS.428.1036M, 2016MNRAS.457.1577G, 2016MNRAS.457..993P}. However, because of the absence of mass/momentum conservation of galaxies, the coupling of two hard modes $k$ to a soft galaxy density perturbation $p$ is not necessarily suppressed by $(p/k)^2$. This leads to additional stochastic contributions which need to be included to provide an accurate description of squeezed-limit $n$-point functions.
\end{itemize}

The squeezed-limit covariance results presented here therefore form only a small subset of a wide range of applications of the response approach, which remains to be explored in the future.

\begin{acknowledgments}

We thank Linda Blot for providing the numerical measurements of the power spectrum covariance, and Daniele Bertolini, Joachim Harnois-D\'eraps, Wayne Hu, and Mikhail Solon for useful discussions.

FS acknowledges support from the Marie Curie Career Integration Grant  (FP7-PEOPLE-2013-CIG) ``FundPhysicsAndLSS,'' and Starting Grant (ERC-2015-STG 678652) ``GrInflaGal'' from the European Research Council.

\end{acknowledgments}

\appendix

\section{Feynman rules for cosmological perturbation theory}\label{app:feynman}

We now spell out the Feynman rules we employ in calculating perturbative predictions for cosmological correlations.  Different conventions are used in the literature, and ours is based on the one used by Ref.~\cite{abolhasani/mirbabayi/pajer:2016}. The rules are as follows: 
\begin{enumerate}
\item An $n$-point correlation function is represented by a collection of diagrams with $n$ outgoing external legs. 
\item Interaction vertices have $m\geq 2$ ingoing lines $\vp_1,\cdots,\vp_m$ coupling to a single outgoing line $\vp$. Each such vertex is assigned a factor
  \be
   m! F_m(\vp_1, \cdots, \vp_m) (2\pi)^3 \d_D(\vp-\vp_{1\cdots m})\,.
  \ee
  In this paper, we assign a positive (negative) sign to outgoing (ingoing) momenta. Each ingoing line has to be directly connected to a propagator (linear power spectrum).\footnote{This is because diagrams that involve interaction vertices directly connected to each other are absorbed into higher-order interaction vertices.}

\item Propagators are represented in our notation as vertices with 2 outgoing lines of equal momentum $k$ as {\includegraphicsbox[scale=0.8]{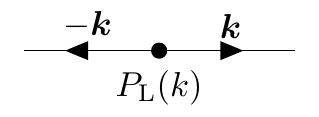}}, and they are assigned a factor $\Plin(k)$. To ease the notation, we often skip labeling these two outgoing lines (which line is which can always be inferred from momentum conservation).

\item All momenta that are not fixed in terms of momentum constraints are integrated over as
  \be
  \int_{\vp} \equiv \int \frac{d^3\vp}{(2\pi)^3}\,.
  \ee
A diagram without any loop integral is said to be a tree-level diagram.
\item Each diagram is multiplied by the {\it symmetry factor}, which accounts for the number of all nonequivalent labelings of external lines and degenerate configurations of the diagram.
\end{enumerate}

In order to include response-type interactions, we augment these rules by one additional rule:

\begin{enumerate}\setcounter{enumi}{5}
\item A second class of interaction vertices is allowed, which have 2 (instead of 1) outgoing lines with momenta $\vk,\vk'$, and $n\geq 1$ incoming lines with momenta $\vp_a$. These vertices are only predictable in the limit where $p = \max\{p_a\} \ll \min\{k,\knl\}$, and $\Sigma_a p_a \ll \min\{k,\knl\}$, but no restriction is placed on the magnitude of the outgoing momenta, which can be fully nonlinear. In this regime, each such vertex is assigned a factor (cf.~Eq.~(\ref{eq:Rndef}))
  \be
 \frac12 \R_n(k; \cdots)\, P_m(k) (2\pi)^3 \d_D(\vk+\vk' - \vp_{1\cdots a})\,.
  \ee
We have not written the arguments of $\R_n$ explicitly here, which are described in detail in \refsec{response}. The factor $1/2$ cancels the trivial permutation $\vk \leftrightarrow \vk'$, which is always present.
\end{enumerate}

\section{Tree-level matter trispectrum}\label{app:tritree}

In this appendix, we explicitly list all the terms that contribute to the $4$-point matter correlation function at tree level. This serves as a concrete example of the discussion in Sec.~\ref{sec:genn} around which types of contributions to the $(n+2)$-point connected correlation function can be represented as power spectrum responses. Explicitly,
\bq\label{eq:tritree}
\<\d(\vk)\d(\vk')\d(\vp_1) \d(\vp_2)\>_c = (2\pi)^3\delta_D\left(\vk+\vk'+\vp_1+\vp_2\right) T(\vk,\vk',\vp_1,\vp_2).
\eq
At tree level, we have that
\bq\label{eq:tritree2}
T^{\rm tree}(\vk,\vk',\vp_1,\vp_2) &=& \Big[4\fii(\vk, -\vk-\vp_1)\fii(\vk', \vk+\vp_1) \Plin(k+p_1)\Plin(k)\Plin(k') \nonumber \\
&& \ \ \ \ \ \ \ \ \ \ \ \ \ \ \ \ \ \ \ \ \ \ \ \ \ \ \ \ \ \ \ \ \ \ \ \ \ \ \ \ \ \ \ \ \ \ \ \ \ \ \ \ \ \ \ \ \ + (11\ {\rm permutations})\Big] \nonumber \\
&&+\left[6\fiii(\vk,\vk',\vp_1)\Plin(k)\Plin(k')\Plin(p_1) + (3\ {\rm permutations})\right] \nonumber \\
&=& T_{\rm a} + T_{\rm b} + T_{\rm c} + T_{\rm d} + T_{\rm \R_2},
\eq
where
\bq
\label{eq:type1} T_{\rm a} &=& 6\fiii(\vk,\vk',\vp_1)\Plin(k)\Plin(k')\Plin(p_1) + (\vp_1 \leftrightarrow \vp_2) \\
\label{eq:type2} T_{\rm b} &=& \big[4\fii(\vk, \vp_1+\vp_2)\fii(\vp_1, -\vp_1-\vp_2) \Plin(p_1+p_2)\Plin(k)\Plin(p_1) + (\vp_1 \leftrightarrow \vp_2) \big] + (\vk \leftrightarrow \vk') \nonumber \\ 
\\
\label{eq:type3} T_{\rm c} &=& \big[4\fii(\vk, -\vk-\vp_2)\fii(\vp_1, \vk+\vp_2) \Plin(k+p_2)\Plin(k)\Plin(p_1) + (\vp_1 \leftrightarrow \vp_2) \big] + (\vk \leftrightarrow \vk') \nonumber \\
\\
\label{eq:type4} T_{\rm d} &=& 4\fii(\vk, -\vk-\vp_1)\fii(\vk', \vk+\vp_1) \Plin(k+p_1)\Plin(k)\Plin(k') + (\vp_1 \leftrightarrow \vp_2) \\
\label{eq:typeR2} T_{\R_2} &=& \big[4\fii(\vp_1, -\vk-\vp_1)\fii(\vp_2, \vk+\vp_1) \Plin(k+p_1)\Plin(p_1)\Plin(p_2) \nonumber \\ 
&& \ \ \ \ \ \ \ \ \ \ \ \ \ \ \ \ \ \ \ \ \ \ \ \ \ \ \ \ \ \ \ \ \ \ \ \ \ \ \ \  + 6\fiii(\vk,\vp_1,\vp_2)\Plin(k)\Plin(p_1)\Plin(p_2) \big]+ (\vk \leftrightarrow \vk') . \nonumber \\
\eq
The terms $T_{\rm a}$, $T_{\rm c}$ and $T_{\rm d}$ belong to type 2 discussed in Sec.~\ref{sec:genn}, and $T_{\rm b}$ is an example of a term that belongs to type 1. As we noted there, this term can be described with the lower order power spectrum response $\R_1$. This can be straightforwardly verified by inserting (cf.~Eqs.~(\ref{eq:bispectree}) and (\ref{eq:Bsq}))
\bq
\R_1^\text{tree}(k, \mu_{\vk,\vp_{12}})\Plin(k) = 2\fii(\vk, \vp_1+\vp_2)\Plin(k) + 2\fii(\vk', \vp_1+\vp_2)\Plin(k')
\eq
into Eq.~(\ref{eq:sqTR1}) at tree level to get Eq.~(\ref{eq:type2}). Finally, the term $T_{\R_2}$ corresponds to the trispectrum terms that are captured by $\R_2$ in the squeezed limit, i.e.,
\bq\label{eq:r2pk_angleapp}
 \< \d(\vk) \d(\vk') \d(\vp_1) \d(\vp_2) \>'_{c,\R_2} &=& \lim_{\{p_a\} \to 0} T_{\R_2}(\vk,\vk',\vp_1,\vp_2) \nonumber \\ 
&\approx& \Bigg[ \mathcal{T}_0 + \mathcal{T}_1\fkone + \mathcal{T}_2\fktwo\Bigg] \Plin(k)\Plin(p_1)\Plin(p_2)\,,
\eq
with
\bq\label{eq:Tterms}
\mathcal{T}_0 &=& \frac{1}{147f_{12}\left(1+f_{12}^2 + 2f_{12}\mu_{12}\right)}\bigg[21f_{12}^4(13+8\mu_1^2)\mu_{12} + 21(13+8\mu_2^2)\mu_{12}\nonumber \\ 
&& \ \ \ \ \ \ \ \ \ \ \ \ \ \ + f_{12}^3\big(628+658\mu_{12}^2 - 280\mu_1\mu_2\mu_{12} + 324\mu_2^2 + 4\mu_1^2\big(95+70\mu_{12}^2+164\mu_2^2\big)\big) \nonumber \\
&& \ \ \ \ \ \ \ \ \ \ \ \ \ \  +2 f_{12}^2\big(112\mu_{12}^3 + 56\mu_1\mu_2 - 336\mu_1\mu_2\mu_{12}^2 + \mu_{12}\big(901 + 408\mu_2^2 + 8\mu_1^2(51+82\mu_2^2)\big)\big) \nonumber \\
&& \ \ \ \ \ \ \ \ \ \ \ \ \ \ + f_{12}\big(628 - 280\mu_1\mu_2\mu_{12} + 380\mu_2^2 + 14\mu_{12}^2(47+20\mu_2^2)\big) + 4\mu_1^2(81+164\mu_2^2)\big)\bigg]  \nonumber \\
\\
\mathcal{T}_1 &=& \frac{-1}{7f_{12}\left(1+f_{12}^2 + 2f_{12}\mu_{12}\right)}\bigg[7f_{12}^4\mu_1^2\mu_{12} + 7\mu_2^2\mu_{12}\nonumber \\ 
&& \ \ \ \ \ \ \ \ \ \ \ \ \ \ \ \ \ \ \ \ \ \ \ \ \ \ \ \ \ \ \ \ \ \ \ \ \ + f_{12}^3\big(-7\mu_1\mu_2\mu_{12} + 13\mu_2^2 + \mu_1^2(16+11\mu_{12}^2 + 23\mu_2^2)\big) \nonumber \\
&& \ \ \ \ \ \ \ \ \ \ \ \ \ \ \ \ \ \ \ \ \ \ \ \ \ \ \ \ \ \ \ \ \ \ \ \ \  + f_{12}^2\big(\mu_1\mu_2(6-20\mu_{12}^2)+33\mu_{12}\mu_2^2 + \mu_1^2\mu_{12}(33+46\mu_2^2)\big) \nonumber \\
&& \ \ \ \ \ \ \ \ \ \ \ \ \ \ \ \ \ \ \ \ \ \ \ \ \ \ \ \ \ \ \ \ \ \ \ \ \ + f_{12}\big(\mu_1^2(13+23\mu_2^2)-7\mu_1\mu_2\mu_{12} + (16+11\mu_{12}^2)\mu_2^2\big)\bigg] \\
\mathcal{T}_2 &=& \mu_1^2\mu_2^2,
\eq
and where we have used the expansions
\bq\label{eq:pkexp2}
\Plin(|\vk+\vp_1|) &=& \Plin(k)\Bigg[1 + \left(\varepsilon_1\mu_1 + \frac{\varepsilon_1^2}{2}\left[1-\mu_1^2\right]\right)\fkone + \frac{\varepsilon_1^2\mu_1^2}{2}\fktwo + \mathcal{O}(\varepsilon_1^3)\Bigg]\\
\Plin(|\vk+\vp_{12}|) &=& \Plin(k)\Bigg[ 1 + \Big(\varepsilon_1\mu_1 + \varepsilon_2\mu_2 + \frac12\big[\varepsilon_1^2(1-\mu_1^2) + \varepsilon_2^2(1-\mu_2^2) \nonumber \\
&&\ \ \ \ \ \ \ \ \ \ \ \ \ \ \ \ \ \ \ \ \ \ \ \ \ \ \ \ \ \ \ \ \ \ \ \  + 2\varepsilon_1\varepsilon_2(\mu_{12} - \mu_1\mu_2)\big]\Big)\fkone \nonumber\\
&& \ \ \ \ \ \ \ \ \ \ \ \ \ \ \ \ \ \ \ \ \ \ \ \ \ \ \ \ \ + \frac12\left(\varepsilon_1^2\mu_1^2 + \varepsilon_2^2\mu_2^2 + 2\varepsilon_1\varepsilon_2\mu_1\mu_2\right)\fktwo + \mathcal{O}(\varepsilon_1^3, \varepsilon_2^3)\Bigg], \nonumber \\
\eq
with $\varepsilon_1=p_1 /k$ and $\varepsilon_2 = p_2/k$. When restricting Eq.~(\ref{eq:r2pk_angleapp}) to the case $f_{12}=1$, the result simplifies considerably and we obtain Eq.~(\ref{eq:r2pk_angle}).


\section{Lagrangian response coefficients}\label{app:RLag}

An equivalent expansion to \refeq{Pkexp} can be performed to yield what we call here the ``Lagrangian'' response coefficients $R_O^L(k)$. These are in some sense the analog of Lagrangian bias parameters for the bias expansion of the matter power spectrum. Consider an operator $O^{[m]}$ whose lowest-order contribution is at $m$-th order. We will drop the indices and contraction with $\hat{\vk}$ here, as it is of no importance for this discussion. At order $m+1$ in perturbation theory, $O^{(m+1)}$ is given by
\be
O^{(m+1)}(\vx) = \sum_{O'^{[m+1]}} Z^{(m+1)}_{O,O'} O'^{(m+1)}(\vx) - s^i \partial_i O^{(m)}(\vx)\,,
\ee
where the sum runs over the same operators that appear in our basis at order $m+1$ (i.e. these operators start at order $m+1$), with coefficients $Z_{O,O'}$ which in the EdS approximation are constants. As an example, consider $O^{[2]} = \d^2$. Then, we have $(\d^2)^{(3)}$ on the left-hand side, and the sum on the right-hand side runs over cubic operators such as $\d^3$. 
The last term is the displacement contribution which, at this order, translates the operator $O^{[m]}$ from the Eulerian to the Lagrangian, or fluid coordinate. A similar division of terms into those which involve the operators in our basis, and those corresponding to the displacement from Eulerian to Lagrangian coordinates, can be made analogously at any order greater than $m+1$. We thus write
\be
O^{(n)} = \sum_{O'^{[n]}} Z^{(n)}_{O,O'} O'^{(n)} + O^{(n)}_{\rm disp}\,.
\label{eq:Oexp}
\ee
Note that $O^{(n)}_{\rm disp}$ only appears for $n \geq m+1$.  The displacement contribution can be rigorously defined as containing those terms which are not invariant under a time-dependent, uniform coordinate transformation, $\vx \to \vx + \v{\xi}(\tau)$. We can now define the Lagrangian response coefficients, by writing the power spectrum expansion at $n$-th order (cf. \refeq{Pkexp2}) as 
\be
\frac{P_m(\vk|\vx)}{P_m(k)}-1 \stackrel{\text{$n$-th order}}{=} \sum_{O^{[n]}} R_O^L(k) O^{(n)}(\vx)
+ \sum_{O^{[m]},m<n} R_O^L(k) \left[O_{\rm disp}^{(n)}\right](\vx)\,.
\label{eq:PkexpL}
\ee
The crucial difference to \refeq{Pkexp2}, and the Eulerian coefficients, is that the $R_O^L$ for operators $O^{[m]},\,m<n$, that start at lower order than $n$, only multiply the displacement part of these lower-order operators. The significance of this expansion is that, when taking the angle-averaged squeezed limit of the equal-time $(n+2)$-point function, the displacement terms cancel by symmetry, as they involve single powers of $\hat{\vp}_a\cdot \hat{\vp}_{b}$. 

Specifically, as shown in Ref.~\cite{response}, after performing an angle average over all soft modes in the squeezed-limit $(n+2)$-point function, a single \emph{isotropic response coefficient} $R_n^L(k)$ remains:
\ba\label{eq:Risodef}
R_n^L(k) \equiv\:& n!  \left[\prod_{a=1}^n\int \frac{d^2\hat{\vp}_a}{4\pi}\right] \R_n\big(k; \{\mu_{\vk,\vp_b}\}, \{\mu_{\vp_b,\vp_c}\}, \{p_b/p_c\}\big) \nonumber \\
=\:& \lim_{\{p_a\}\to 0}\; \frac{1}{P_m(k) \Plin(p_1) \cdots \Plin(p_{n})} \left[\prod_{a=1}^n\int \frac{d^2\hat{\vp}_a}{4\pi}\right] \< \d(\vk) \d(\vk') \d(\vp_1) \cdots \d(\vp_{n}) \>'_{c,\R_n}\,,
\ea
where the second equality is obtained using Eq.~(\ref{eq:sqnpt}). The second relation has in fact been derived in Ref.~\cite{response}, who however did not emphasize that only the $\R_n$-type contributions as written in Eq.~(\ref{eq:sqnpt}) are to be included in the $(n+2)$-point connected correlator (that is, not the contributions which involve lower $\R_m$ and PT couplings between soft modes). Nevertheless, Eq.~(\ref{eq:Risodef}) establishes a well-defined relation between the angle-averaged squeezed $(n+2)$-point function and the $n$-th order isotropic Lagrangian response coefficient $R_n^L(k)$. 

Crucially, both $R_O^E$ and $R_O^L$ are physical coefficients, which moreover can be related to each other unambiguously in perturbation theory. For this, we separate \refeq{Pkexp2} into operators which start at $n$-th order, and those which start at lower order. For the latter, we then insert \refeq{Oexp}. In this matching, we only need to consider the boost-invariant (non-displacement) terms.  We then obtain
\ba
\sum_{O^{[n]}} R_O^E(k)  O^{(n)}(\vx) + \sum_{O^{[m]},m<n} \sum_{O'^{[n]}} Z^{(n)}_{O,O'} R_{O'}^E(k) O^{\prime (n)}(\vx)  = \sum_{O^{[n]}} R_O^L(k) O^{(n)}(\vx)
\,,
\ea
and hence 
\be
R_O^L(k) = R_O^E(k) + \sum_{O'^{[n]}} Z_{O',O}^{(n)} R_{O'}^E(k)
\label{eq:RLRE}
\ee
where the sum runs over all lower-order operators in whose $n$-th order expression in perturbation theory the $n$-th order operator $O$ appears.

For completeness, we now give the explicit expressions for the Lagrangian response coefficients in \refeq{PkexpL} at second order. From \refeq{RLRE}, it follows that the first-order responses $R_1$ and $R_K$ are identical in both Eulerian and Lagrangian expansions. The only difference to \refeq{Pkexp} is that the contributions from nonlinear evolution of the first-order operators disappear. Instead, we only explicitly include terms that displace the first-order terms from the Eulerian  (with respect to which the long-wavelength fields $\d(\vp_1)$ and $\d(\vp_2)$ are defined) to the Lagrangian position, specifically $-s_{(1)}^k \partial_k K_{ij}^{(1)}$ and $-s_{(1)}^k \partial_k \d^{(1)}$. Then, \refeq{PR2} is replaced with 
\bq\label{eq:PR2L}
\frac{P_m\left(\vk| \d(\vp_1) \d(\vp_2) \right)}{P_m(k)}-1 &=& R^L_1(k) \frac{\mu_{12}}2 \bigg[ \frac{p_1}{p_2} +  \frac{p_2}{p_1}  \bigg]\d(\vp_1)\d(\vp_2) \nonumber \\ 
&&+ R^L_K(k) \frac{\mu_{12}}2 \bigg[ \left(\mu_1^2-\frac13\right) \frac{p_1}{p_2} + \left(\mu_2^2-\frac13\right) \frac{p_2}{p_1} \bigg] \nonumber \\
&&+ \frac12 R^L_2(k) \bigg[\d(\vp_1)\d(\vp_2)\bigg] + R^L_{K \d}(k) \bigg[\khat^i \khat^j K_{ij}(\vp_1) \d(\vp_2)\bigg] \nonumber \\ 
&&+ R^L_{K^2}(k)\bigg[ K_{ij}(\vp_1) K^{ij}(\vp_2)\bigg] + R^L_{K.K}(k) \bigg[\khat^i \khat^j K_{ik}(\vp_1) K^k_{\  j}(\vp_2) \bigg] \nonumber \\
&&+ R^L_{KK}(k) \bigg[\khat^j \khat^j \khat^l \khat^m K_{ij}(\vp_1) K_{lm}(\vp_2)\bigg] + R^L_{\Ote}(k) \bigg[\khat^i \khat^j \Ote_{ij}(\vp_1,\vp_2)\bigg]. \nonumber \\
\eq
The explicit kinematic dependence of the Lagrangian response \refeq{PR2L} becomes, for $p_1=p_2$,
\ba
\label{eq:PR2L_angle}
\R^L_2(k, \mu_1,\mu_2,\mu_{12}, f_{12}=1) =\:& 
R^L_1(k) \mu_{12} + R^L_K(k)\Bigg[ \frac12  \mu_{12} \left( \mu_1^2 + \mu_2^2 - \frac23 \right) \Bigg] + \frac12 R^L_2(k) \vs
& + \frac12 R^L_{K \d}(k) \Bigg[\mu_1^2 + \mu_2^2 - \frac23 \Bigg] + R^L_{K^2}(k) \Bigg[\mu_{12}^2 - \frac13\Bigg] \nonumber \\
& +  R^L_{K.K}(k) \Bigg[\mu_1 \mu_2 \mu_{12} {-\frac13\mu_1^2-\frac13\mu_2^2+\frac19}\Bigg] \nonumber \\
& + R^L_{KK}(k) \Bigg[\mu_1^2\mu_2^2 - \frac13\left(\mu_1^2 + \mu_2^2\right) + \frac19\Bigg] \nonumber \\
& + {\frac32} R^L_{\Ote}(k) \Bigg[ \frac12 (\mu_1 + \mu_2)^2 (1 - \mu_{12})  - \frac13 (1 - \mu_{12}^2)\Bigg]\,.
\ea
\begin{figure}
	\centering
	\includegraphics[width=\textwidth]{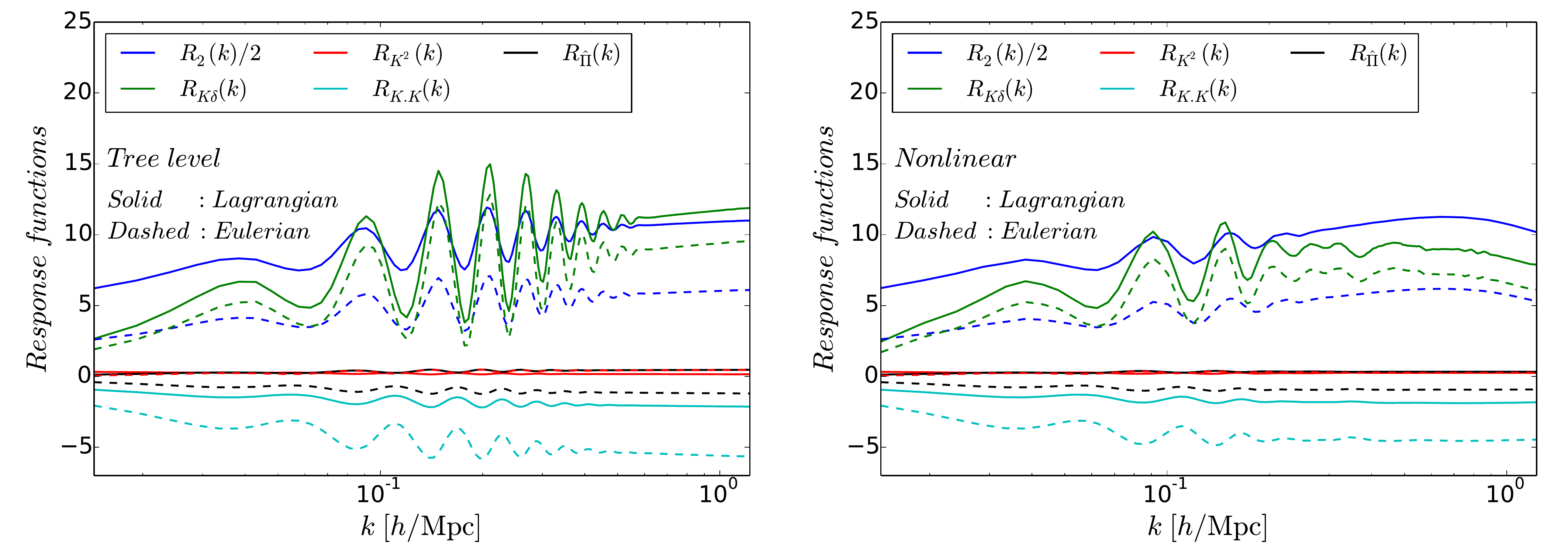}
	\caption{Comparison between the Lagrangian and Eulerian response coefficients $R_O(k)$, as labeled. The functions $R_1$, $R_K$ and $R_{KK}$ have the same Lagrangian and Eulerian expressions and are not shown (cf.~Fig.~\ref{fig:rs}). The left and right panels show the tree-level and the nonlinear results, respectively. The nonlinear extrapolation of the $R_O^L$ is obtained from the extrapolation of the $R_O^E$ using Eqs.~(\ref{eq:rsEL}).}
\label{fig:rsevl}
\end{figure}
By matching to the tree-level trispectrum as in the case of the Eulerian expansion, the Lagrangian response coefficients are obtained as
\bq\label{eq:rsL}
R^L_2(k) &=& \frac{8420}{1323} - \frac{100}{63}\fkone + \frac19\fktwo,\ \ ; \ \ \ R^L_{K\delta}(k) = \frac{1348}{441} - \frac{55}{21}\fkone + \frac13\fktwo, \nonumber \\
R^L_{K^2}(k) &=& \frac{20}{63} + \frac{1}{14}\fkone,\ \ \ \ \ \ \ \ \ \ \ \ \ \ \ \ \ \ \ \ \ \ \ \ ; \ R^L_{K.K}(k) = -\frac{20}{21} +\frac12 \fkone, \nonumber \\
R^L_{KK}(k) &=& R^E_{KK}(k) \ \ \ \ \ \ \ \ \ \ \ \ \ \ \ \ \ \ \ \ \ \ \ \ \ \ \ \ \ \ \ \ \ \ \ \ ; \ \ \ \ R^L_{\Ote}(k) = \frac{8}{63} -\frac17\fkone\,,
\eq
where we have dropped the superscript ``tree'' for clarity, and did not repeat the trivially identical results for $R_1$ and $R_K$ (note however that $R^L_{KK} = R^E_{KK}$, as well). The Eulerian and Lagrangian coefficients can therefore be related as
\bq\label{eq:rsEL}
R^{L}_2 &=&\frac{34}{21}R^E_1(k) + R^E_2(k), \nonumber \\
R^{L}_{K\delta} &=&\frac{2}{3}R^E_K(k) + R^E_{K\delta}(k), \nonumber \\
R^{L}_{K^2} &=&\frac{2}{7}R^E_1(k) - \frac{1}{3}R^E_K(k)+ R^E_{K^2}(k), \nonumber \\
R^{L}_{K.K} &=&R^E_K(k)+ R^E_{K.K}(k), \nonumber \\
R^{L}_{\Ote} &=&\frac{10}{21}R^E_K(k)+ R^E_{\Ote}(k) .
\eq
This relation in fact is \emph{not} restricted to the tree-level responses, but also holds for the fully nonlinear response coefficients, as derived above. In fact, the first relation was already obtained by Ref.~\cite{response}.  Using that
\ba
\d^{(2)} =\:& \frac{17}{21} \d^2 + \frac27 (K_{ij})^2 - s^i \partial_i \d \vs
K_{ij}^{(2)} =\:& \frac{10}{21} \Ote_{ij} + K_{ik} K^k_{\  j} - \frac13 \d_{ij} (K_{kl})^2 + \frac23 K_{ij}\d - s^k \partial_k K_{ij}\,,
\ea
where all quantities on the right-hand sides are evaluated at linear order, it is straightforward to verify that \refeq{RLRE} indeed yields all relations in \refeq{rsEL}. 

Figure \ref{fig:rsevl} compares the second-order Eulerian and Lagrangian response coefficients that are different. We see that, in most cases, the Eulerian coefficients are smaller than the Lagrangian ones. This is because the Lagrangian coefficients contain some of the coupling between long-wavelength modes that is not included in the Eulerian coefficients, as has been already noticed in the case of $R_2$ and $R_3$ in Ref.~\cite{response}.

\bibliography{REFS}

\end{document}